\documentclass[aps,prc,superscriptaddress,twoside,twocolumn,nofootinbib,showpacs]{revtex4}
\usepackage{amsmath,amssymb}
\usepackage{mathtools}
\usepackage{bbold}
\usepackage[usenames, dvipsnames]{xcolor}
\usepackage{braket}
\usepackage{graphicx,bm}
\usepackage{slashed}
\usepackage{booktabs}
\usepackage{tensor}
\usepackage{textgreek}

\usepackage{multirow}

\usepackage{changes}

\allowdisplaybreaks

\newcommand{\mbk}{\mathbf{k}}
\newcommand{\mbkp}{\mathbf{k}^{\prime}}
\newcommand{\psig}{P_\mathbf{\sigma}}
\newcommand{\mbkpp}{\mathbf{k}^{\prime 2}}
\newcommand{\mbsig}{\boldsymbol{\sigma}}
\newcommand{\rij}{\mathbf{r}_i -\mathbf{r}_j}
\newcommand{\mbr}{\mathbf{r}}

\newcommand{\bnabla}{\nabla}

\newcommand{\rhon}{\rho_n}
\newcommand{\rhop}{\rho_p}

\newcommand{\mc}{\mathcal}

\newcommand{\pt}{\partial}

\newcommand{\vare}{\varepsilon}
\newcommand{\taun}{\tau_{n}}
\newcommand{\taup}{\tau_{p}}

\newcommand{\mbrp}{\mathbf{r}^{\prime}}

\newcommand{\mb}{\mathbf}

\newcommand{\mun}{\mu_{n}}
\newcommand{\mup}{\mu_{p}}
\newcommand{\msun}{M_{\odot}}

\begin{document}


\title{\boldmath Neutron star crusts from mean field models constrained \\ by chiral effective field theory}
\date{\today}

\author{Yeunhwan \surname{Lim} }
\email{ylim@tamu.edu}
\affiliation{Cyclotron Institute, Texas A\&M University, College Station, TX 77843, USA}

\author{Jeremy W. \surname{Holt} }
\email{holt@physics.tamu.edu}
\affiliation{Cyclotron Institute, Texas A\&M University, College Station, TX 77843, USA}
\affiliation{Department of Physics and Astronomy, Texas A\&M University, College Station, TX 77843, USA}

\begin{abstract}
We investigate the structure of neutron star crusts, including the crust-core boundary, based on
new Skyrme mean field models constrained by the bulk-matter equation of state from 
chiral effective field theory and the ground-state energies of doubly-magic nuclei. Nuclear pasta 
phases are studied using both the liquid drop model as well as the Thomas-Fermi approximation.
We compare the energy per nucleon for each geometry (spherical nuclei, cylindrical nuclei, 
nuclear slabs, cylindrical holes, and spherical holes) to obtain the ground state phase
as a function of density.
We find that the size of the Wigner-Seitz cell depends strongly on the model parameters, 
especially the coefficients of the density gradient interaction terms.
We employ also the thermodynamic instability method to check the validity of the numerical 
solutions based on energy comparisons.
\end{abstract}


\maketitle

\section{Introduction}
Neutron stars offer the possibility to study matter under extreme conditions (in 
density and neutron-to-proton ratio) inaccessible to laboratory experiments on Earth. 
The inner core of a neutron star may reach densities as high as five to ten times 
nuclear saturation density, 
a regime for which no well-converged theoretical expansions 
are presently available. The structure and composition of the inner core is
consequently highly uncertain and 
may contain deconfined quark matter~\cite{Weber2005193,0004-637X-629-2-969,Weissenborn:2011qu}, 
hyperonic matter~\cite{SchaffnerBielich:2000yj,Weissenborn:2011kb,Weissenborn:2011ut,Lim:2014sra,Chatterjee2016}, 
or meson condensates~\cite{Baym:1973zk,Thorsson:1993bu,Glendenning:1998zx,Lim:2013tqa}.
In contrast, the inner crust and outer core span densities 
from $n \simeq 4 \times 10^{11} - 5 \times 10^{14}$\,g/cm$^3$, corresponding to 
nucleon Fermi momenta of $k_F \lesssim 400$\,MeV, which is much less than the chiral
symmetry breaking scale of $\Lambda_\chi = 4 \pi f_\pi \simeq 1$\,GeV. Chiral effective
field theory (EFT) \cite{weinberg79} may therefore provide a suitable theoretical framework 
for exploring neutron star matter at these densities.

In recent years there has been significant progress in the development of realistic chiral 
nucleon-nucleon (NN) forces \cite{epelbaum09rmp,machleidt11,epelbaum15,entem15} at 
and beyond next-to-next-to-next-to-leading order (N3LO) in the chiral power counting. Nuclear
many-body forces become relevant in homogeneous matter at densities larger than  
$n \gtrsim 0.25 n_0$ (where $n_0 = 2.4 \times 10^{14}$\,g/cm$^3$ is the
saturation density of nuclear matter) and have been included in numerous studies 
of the cold nuclear and neutron matter equations of state (EOS)
\cite{epelbaum09epja,hebeler10,hebeler11,coraggio13,gezerlis13,tews13,coraggio14,roggero14,wlazlowski14,carbone14,tews16}. Neutron star structure and evolution
requires in addition the equation of state at arbitrary isospin-asymmetry \cite{drischler14,wellenhofer16} 
and finite temperature \cite{tolos08,wellenhofer14,wellenhofer15}, which has been computed
consistently with the same chiral nuclear force models and many-body methods.
The inhomogeneous phase of nuclear matter encountered in neutron star crusts 
depends also on gradient contributions to the energy density. Previous work has focused on the
leading-order Hartree-Fock contribution to the isoscalar and isovector gradient couplings from
the density matrix expansion \cite{bogner08,holt11epj,kaiser12}, ab initio
studies of the isovector gradient coupling strength from quantum Monte Carlo simulations of
pure neutron matter \cite{Gandolfi:2010za,Buraczynski:2015oaa}, and nuclear response functions
in Fermi liquid theory \cite{lykasov08,davesne15,holt12,holt13prc}.

Neutron star crusts have been studied using phenomenological
liquid drop models~\cite{Ravenhall:1983uh, Douchin:2001sv, 0067-0049-204-1-9} and the
Thomas-Fermi approximation~\cite{Oyamatsu:1993zz,PhysRevC.88.025801}.
Nuclear pasta phases resulting from the competition
between the Coulomb interaction and nuclear surface tension
were also treated in the liquid drop and Thomas-Fermi methods.
More sophisticated approaches to the nuclear pasta phase have been investigated
using the Skyrme-Hartree Fock approximation~\cite{PhysRevC.79.055801,PhysRevLett.109.151101,PhysRevC.92.045806}
and molecular dynamic simulations~\cite{PhysRevC.72.035801,PhysRevC.77.035806,PhysRevC.88.065807}.

In the present work we utilize recent results for the homogeneous nuclear matter equation
of state from chiral EFT to develop new Skyrme mean field parametrizations that enable the study of finite 
nuclei, inhomogeneous nuclear matter in neutron star crusts, and the mass-radius relation of neutron 
stars. Recent works \cite{Brown:2013pwa,Bulgac:2015sba,rrapaj16} have used the low-density equation of
state of neutron matter from chiral EFT to constrain nonrelativistic and relativistic mean field models, 
while the present study includes the full isospin-asymmetric 
matter equation of state at second order in perturbation theory up to $n = 2 n_0$ as a fitting 
constraint. Several chiral nuclear force models are considered in order to estimate the
theoretical uncertainty.

We find that the traditional Skyrme model cannot accommodate the density dependence of the
nuclear equations of state derived from chiral effective field theory. We therefore introduce additional 
interaction terms in the Skyrme Hamiltonian that go as the next higher power of the Fermi momentum. 
This enables an accurate reproduction of the bulk-matter equation of state from chiral EFT.
Using the new models, we investigate the phase of sub-saturation
nuclear matter, which is expected to be present at the boundary
between the outer core and inner crust of neutron stars, an environment that
is highly neutron rich. Indeed the proton fraction
of nuclear matter in beta equilibrium at the crust-core boundary is roughly
$\sim 3\%$. 
In the boundary region, nuclear matter experiences a shape change
caused by the competition between the repulsive Coulomb interaction and surface tension.
We adopt the analytic solution of the Coulomb interaction
in discrete dimensions to study the phase of nuclear matter in the liquid 
drop model (LDM) formalism.
The energy per nucleon of nuclear matter
determines the lowest energy state and therefore the discrete shape in the pasta phase.
We also study inhomogeneous nuclear matter by employing the Thomas-Fermi (TF)
approximation employing a parameterized density profile (PDP) for
neutrons and protons. 

The paper is organized as follows. In Section \ref{sec:model} we describe the Skyrme force 
model used to investigate the neutron star inner crust and outer core. 
The traditional Skyrme model is extended in order to reproduce the homogeneous matter
equation of state of isospin-asymmetric nuclear matter from chiral effective field theory as well
as the ground state energies of doubly magic nuclei.
In Section \ref{sec:boundary}, we present the numerical method to determine the transition
density for the core-crust boundary. The liquid drop model, Thomas-Fermi approximation, and
thermodynamic instability methods are then employed to find the transition densities.
We summarize our results in Section \ref{sec:con}.

\section{Nuclear Model}
\label{sec:model}

We begin by describing the microscopic chiral nuclear force models \cite{entem03,coraggio14} 
employed in the present study. The two-body force is treated at N3LO in the chiral expansion,
and the 24 low-energy constants associated with NN contact terms are fitted to elastic 
nucleon-nucleon scattering phase shifts and deuteron properties. The three-body force is 
treated at N2LO, and the $c_E$ and $c_D$ low-energy constants associated with the contact
three-body force and one-pion exchange three-body force, respectively, are fitted to reproduce 
the ground-state energies of $^3$H and $^3$He as well as the beta-decay lifetime of $^3$H.
The resolution scale is set by the momentum-space cutoff $\Lambda$, which is varied over the
range $414\,{\rm MeV} < \Lambda < 500$\,MeV. At this resolution scale many-body perturbation
theory is well converged, and the resulting neutron matter equation of state below saturation
density is strongly constrained \cite{Holt:2016pjb}. Cutoff variation provides only one means to 
study the theoretical uncertainties in chiral effective field theory, and future work will be devoted
understanding better the errors due to neglected higher-order terms in the chiral expansion. 

To be specific we use three different values of the momentum-space cutoff 
$\Lambda=414$, $450$, $500$~MeV and denote the corresponding nuclear potentials
as n3lo414, n3lo450, and n3lo500. The strategy is then to identify what approximations 
are needed in each case to provide an accurate description of the bulk matter
equation of state in the vicinity of nuclear matter saturation. As shown in previous work
\cite{coraggio14}, the chiral potentials with the two 
lowest cutoff values give reasonable nuclear matter properties at second-order in 
many-body perturbation theory with Hartree-Fock intermediate-state propagators.
In particular, the saturation energy lies in the range $E/A = -(15.7 - 16.2)$\,MeV while the 
saturation density lies in the range $n_0 = (0.165-0.174)$\,fm$^{-3}$. At the same 
approximation in many-body perturbation theory, the $\Lambda = 500$\,MeV
chiral potential exhibits too little attraction, and the binding energy per nucleon at saturation 
density is only $E/A \simeq -11.5$\,MeV. We therefore employ for this potential 
second-order perturbation theory with free-particle intermediate-state energies, which on
the one hand accounts for theory uncertainties associated with the choice of the single-particle 
energy spectrum and on the other hand leads to an improved description of nuclear matter 
saturation. The latter results from a larger density of states
near the Fermi surface that enhances the overall attraction from the second-order 
perturbative contribution. In this case the saturation energy and density are $E/A = -15.9$\,MeV
and $n_0 = 0.171$\,fm$^{-3}$, respectively. 

The calculations outlined above have been extended in the present work to describe cold 
nuclear matter at arbitrary isospin asymmetry. The resulting equations of state are then used 
as data in fitting new Skyrme model parametrizations. In addition, the density-gradient 
contributions to the nuclear energy density, which have important effects on the structure of
the neutron star inner crust, are constrained by including the ground-state energies of 
doubly-magic nuclei in the $\chi^2$ minimization function for the Skyrme model parameters.

The same two- and three-body chiral potentials have also been used in numerous
studies of nuclear dynamics and thermodynamics (for recent reviews, see Refs.\
\cite{holt13ppnp,holt16pr}. In particular, the critical endpoint of the first-order liquid-gas
phase transition line was found \cite{wellenhofer14} to be consistent with recent empirical 
determinations \cite{elliott13}, and the low-density--high-temperature equation of state of pure 
neutron matter was found \cite{wellenhofer15} to be in very good agreement with the 
model-independent virial expansion. The applications described below focus on the cold neutron
star composition and equation of state, but we may anticipate future extensions to finite temperature
matter employing a strategy similar to that described above.

The energy density in dense nuclear matter can be expanded in powers of the 
proton and neutron Fermi momenta, $k_f^p = (3 \pi^2 n_p)^{1/3}$ and
$k_f^n = (3 \pi^2 n_n)^{1/3}$, as follows
\begin{equation}
\begin{aligned}
\vare = & \frac{\hbar^2}{2m}(\tau_n + \taup) 
+ \alpha_L(n_n^2 + n_p^2) + 2\alpha_U n_n n_p \\
& + \left[ \eta_L(n_n^2 + n_p^2) + 2\eta_Un_n n_p \right]n^{\gamma}
\,,
\end{aligned}
\label{eq:pol}
\end{equation}
where 
\begin{equation}
\begin{aligned}
\taun=& \frac{3}{5}(3\pi^2)^{2/3}n^{5/3}(1-x)^{5/3}, \\
\taup=& \frac{3}{5}(3\pi^2)^{2/3}n^{5/3}(1-x)^{5/3},
\end{aligned}
\end{equation}
$n_n = n(1-x)$, and $n_p=nx$.
The above approximation can explain $\chi$EFT asymmetric matter results
quite well with small deviation ($<1\%$), at least for $T=0$~MeV. However, it 
cannot be used to calculate the properties of finite nuclei directly unless we
find the surface tension in the liquid drop model or the gradient terms 
in the Thomas-Fermi approximation

The polynomial expansion in Eq.~\eqref{eq:pol} can be derived from the
phenomenological Skyrme nucleon-nucleon interaction, given by
\begin{equation}
\begin{aligned}
v_{i,j} &({\mb r}_i,{\mb r}_j) =  t_0(1+x_0\psig)\delta(\rij) \\
& +\frac{t_1}{2}(1+x_1\psig) 
 \Bigl[ \delta(\rij)\mbk^2  + \mbkpp\delta(\rij) \Bigr] \\
& + t_2(1+ x_2\psig)\mbkp \cdot \delta (\rij)\mbk  \\
&+ \frac{1}{6}t_3(1+x_3\psig)\rho^{\alpha}\delta(\rij) \\
& + iW_0 \mbkp \delta(\rij)\times \mbk
\cdot (\mbsig_i+\mbsig_j)
\,,
\end{aligned}
\label{eq:skyint}
\end{equation} 
where $P_\sigma$ is the spin exchange operator,
the local density $\rho$ is evaluated at $(\mathbf{r}_i+\mathbf{r}_j)/2$,
$\mbk =\frac{1}{2i}(\bnabla_{i}-\bnabla_j)$, and
$\mbkp = -\frac{1}{2i}(\bnabla^{\prime}_i -\bnabla^{\prime}_j)$.
\\

\begin{table}[t]
\caption{Skyrme force parameters fitted to the chiral N3LO asymmetric matter 
equation of state and finite nuclei binding energies. The parameters have units such 
that the energy density is given in MeV fm$^{-3}$.}
\begin{tabular}{ccccc}
\hline
\hline
   & ~Sk$\chi$414~        & ~Sk$\chi$450~  & ~Sk$\chi$500~ \\
\hline    
$t_0$      &  ~$-1734.0261 $~ &  ~$-1803.2928$~    &  ~$-1747.48258 $~ \\
$t_1$      &  $255.6550 $     &  $  301.8208$      &  $241.31968 $     \\
$t_2$      &  $-264.0678 $    &  $-273.2827$       &  $-331.04118 $    \\
$t_3$      &  $12219.5884 $   &  $12783.8619$      &  $12491.50533 $   \\
$t_4$      &  $556.1320 $     &  $564.1049$        &  $405.03174 $     \\
$x_0$      &  $0.4679 $       &  $0.4430$          &  $0.59530 $       \\
$x_1$      &  $-0.5756 $      &  $-0.3622$         &  $-1.15893 $      \\
$x_2$      &  $-0.3955 $      &  $-0.4105$         &  $ -0.58432$      \\
$x_3$      &  $0.7687 $       &  $ 0.6545$         &  $ 1.20050$       \\
$x_4$      &  $-15.8761 $     &  $-11.3160$        &  $ -25.49381$     \\
$\gamma_1$ &  $1/3$           &  $1/3$             &  $1/3$            \\
$\gamma_2$ &  $1$             &  $1$               &  $1$              \\
$W_0$      &  $93.7236 $      &  $106.4288$        &  $98.08897$       \\
\hline
\end{tabular}
\label{tb:skyrme_eft}
\end{table}

Traditional Skyrme force models have 10 parameters which can be
fitted to the binding energies of finite nuclei, neutron skin thicknesses,
bulk matter properties, and neutron matter calculations.
However, we find that this number of parameters is insufficient to
reflect both the equation of state of asymmetric nuclear matter from
chiral EFT as well as the properties of finite nuclei.
We therefore extend the traditional Skyrme force model by adding
extra density dependent terms of the form
\begin{equation}
v_{ij} \rightarrow v_{ij} + \frac{1}{6}t_4(1+x_4P_\sigma)
\rho^{\gamma_2}\delta(\mbr_i -\mbr_j)\,.
\end{equation}
We determine the Skyrme Hartree-Fock parameters from fitting to the
recent $\chi$EFT asymmetric nuclear matter calculations outlined in Ref.\ 
\cite{wellenhofer16} together with the binding energies of 
doubly closed shell nuclei. We define the $\chi^2$ minimization function:
\begin{equation}
\begin{aligned}
& \chi^2(x_0,\dots,x_4,t_0,\dots,t_4,W_0) \\
& \quad =  w_b\left[ \frac{1}{N_iN_j}
                 \sum
                 \left\{ 
                 \frac{\mathcal{E}^{\mathrm{EFT}}(n_i,x_j)-\mathcal{E}^{\mathrm{Sk}}(n_i,x_j)}
                 {\mathrm{MeV}}\right\}^2\right] \\
&\quad   + w_{n_0}(0.16-\rho_0~\mathrm{fm}^3)^2 
       + w_B(-16+B^{\mathrm{Sk}}~\mathrm{MeV}^{-1})^2 \\
&\quad    + w_{F}\left[\frac{1}{N_k}\sum
\left( \frac{B_k^{\mathrm{Exp.}}-B_k^{\mathrm{Sk}}}{\mathrm{MeV}}\right)^2\right]\,
\end{aligned}
\label{fit}
\end{equation}
with weighting factors \{$w_b$, $w_{n_0}$, $w_B$, $w_F$\}.

Since Hartree-Fock theory is the lowest order approximation in a systematic
many-body perturbation theory expansion, there is no clean one-to-one 
correspondence between the Skyrme parameters and the chiral expansion 
coefficients. It is, however, possible to reproduce properties of the chiral EFT 
equation of state from a simplified Skyrme mean field model. The desirable 
aspect of the Skyrme parametrization is that it enable us to then calculate also the
properties of finite nuclei, such as their density profiles and binding energies,
as well as the composition and structure of neutron star inner crusts.

\begin{figure}[t]
\includegraphics[scale=0.45]{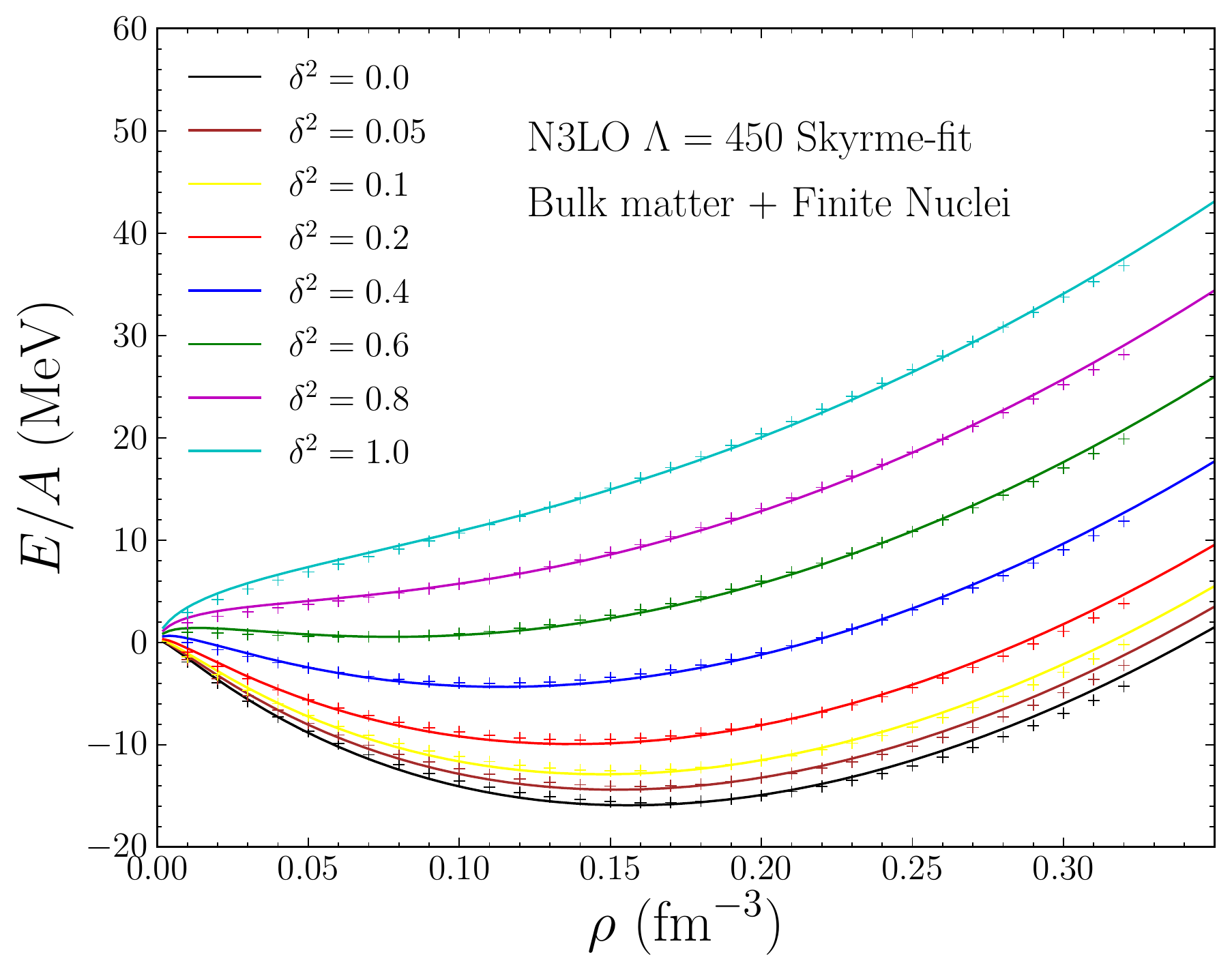}
\caption{(Color online) Comparison of the energy per baryon in asymmetric nuclear matter 
from chiral EFT ($\Lambda=450$\,MeV) and its Skyrme fitting model. The isospin
asymmetry is denoted by $\delta = (n_n - n_p)/(n_n+n_p)$.}
\label{fig:epa_eft}
\end{figure}

We present the new Skyrme parametrizations in Table \ref{tb:skyrme_eft}. We
set $\gamma_1 =1/3$ and $\gamma_2 =1$ in all cases.
This can be justified when we consider that the energy density of bulk nuclear matter
can be expanded as a function of the Fermi momentum $k_f$. 
Note that $x_4$ is much larger than the other $x$'s in the parametrization. 
This indicates that spin exchange interactions give very large attraction in 
dense matter within the extended Skyrme formalism.
Figure \ref{fig:epa_eft} shows the energy per baryon in asymmetric nuclear matter
from both chiral effective field theory and Skyrme phenomenology. 
The `$+$' denotes the energy per baryon from $\chi$EFT with $\Lambda=450$\,MeV,
while the solid lines are results from the new Skyrme models derived in our work. 
The deviations get larger as the total baryon number density increases, but overall
the agreement is quite satisfactory given the simplicity of the Skyrme mean field
model. We have performed the same fitting procedure also for the $\Lambda=414$\,MeV
and $\Lambda=500$\,MeV chiral nuclear potentials, and in these cases the fit is of the 
same quality as that shown in Fig.\ \ref{fig:epa_eft} for the case $\Lambda=450$\,MeV.
We include as well the total binding energy of doubly magic nuclei in the 
$\chi^2$ minimization function for the Skyrme parametrizations.
Table \ref{tb:fnuc} shows the results of the Skyrme Hartree-Fock calculations 
compared to the experimental values \cite{Audi:2014wda}.
\begin{table}[t]
\caption{Skyrme Hartree-Fock results for the binding energies (in units of MeV) 
of doubly closed shell nuclei together with bulk nuclear matter properties \cite{Dutra2012}.}
\label{tb:fnuc}
\begin{tabular}{lccccccc}
\hline
\hline
            & Exp.      & Sk$\chi$414              & Sk$\chi$450              & Sk$\chi$500       \\
\hline         
$^{16}$O    & 127.62    & 126.73            & 126.93            &   127.07  \\
$^{40}$Ca   & 342.05    & 342.63            & 341.93            &   341.43  \\  
$^{48}$Ca   & 415.99    & 416.66            & 416.69            &   417.24  \\ 
$^{56}$Ni   & 483.99    & 482.29            & 482.32            &   482.38  \\ 
$^{100}$Sn  & 825.78    & 826.20            & 825.69            &   822.55  \\
$^{132}$Sn  & 1102.90   & 1103.05           & 1103.22           &   1106.91 \\ 
$^{208}$Pb  & 1636.44   & 1635.88           & 1636.21           &   1635.30 \\
\hline
\hline
$\rho_0$ ($\mathrm{fm}^{-3}$) &   $0.160 \pm 0.005$  &   0.1697          &   0.1562             &  0.1679  \\
$B$ (MeV)                     &   $16.0 \pm 0.5$     &   16.1987         & 15.9262              &  15.9895 \\  
$K$ (MeV)                     &   $230 \pm 30$       &  243.19           & 239.53               &  238.16  \\
$S_v$ (MeV)                   &   $32.5\pm 2.5$      &  32.3456          & 30.6346              &  29.1167 \\ 
$L$ (MeV)                     &   $58 \pm 18$    &  51.9307          &  42.0518             &  40.7415 \\
\hline
\end{tabular}
\end{table}

Having determined all Skyrme model parameters from the $\chi^2$ fitting function
in Eq.\ (\ref{fit}), we now check theoretical predictions for bulk matter and finite nuclei. 
Also in Table \ref{tb:fnuc} we show the properties of nuclear matter around the saturation 
density, including the saturation energy per particle $B$, the nuclear incompressibility
$K$, the isospin-asymmetry energy $S_v$, and the isospin-asymmetry slope 
parameter $L$. Overall the microscopic predictions agree very favorably with experimental 
constraints \cite{Dutra2012}.

\begin{figure}
\includegraphics[scale=0.45]{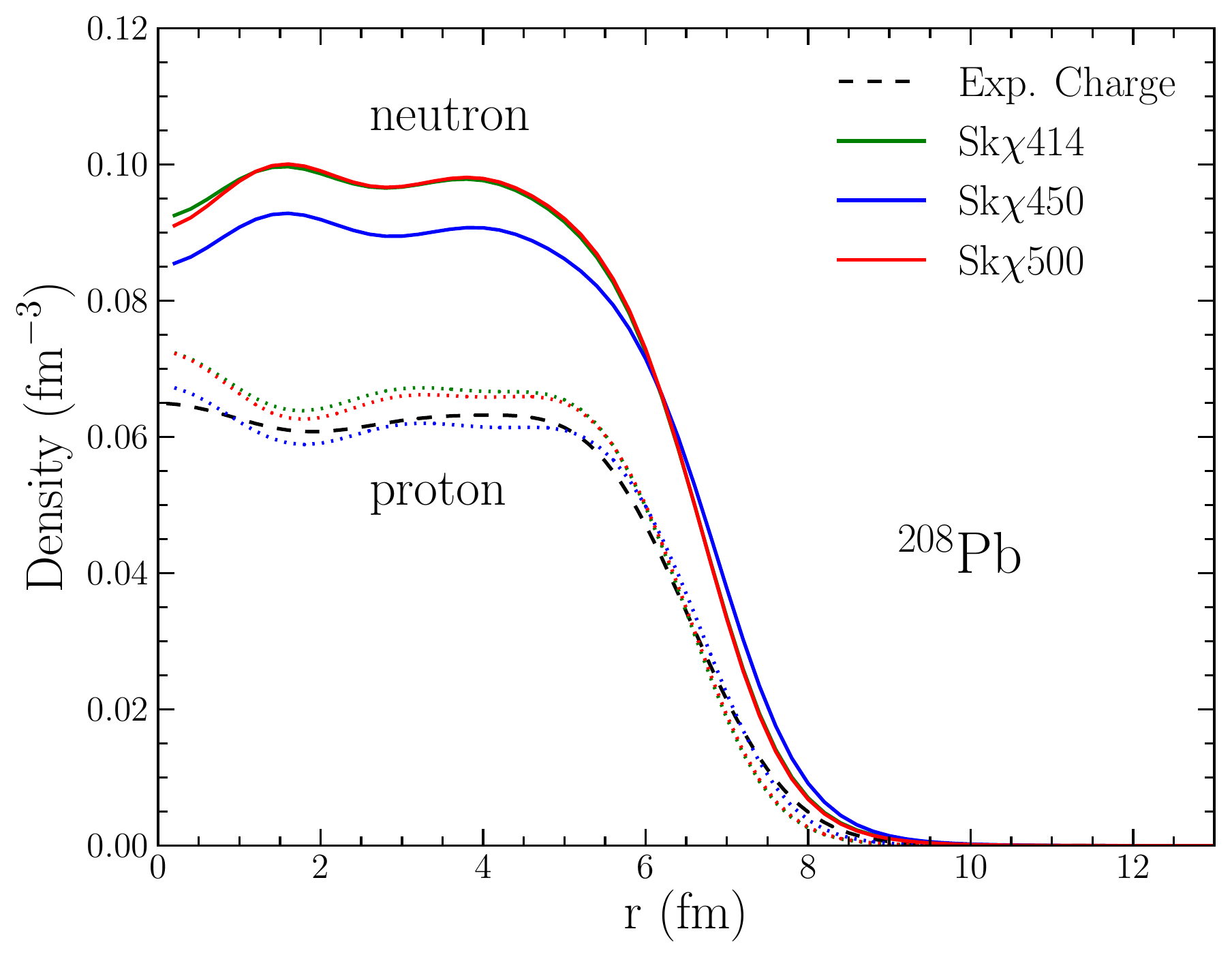}
\caption{(Color online) Density profile of neutron and proton in $^{208}$Pb using Skyrme Hartree Fock.
Experimental charge density is also added for comparison.}
\label{fig:pb}
\end{figure}

As an example of the Skyrme Hartree-Fock calculations for finite nuclei, 
we present the density profile of $^{208}$Pb in Fig.\ \ref{fig:pb}. 
The experimental charge density~\cite{DeJager:1987qc} is included for comparison. 
The central density of $^{208}$Pb from the $\Lambda=414$\,MeV and 
$\Lambda=500$\,MeV chiral potentials is greater than that from the
$\Lambda=450$\,MeV potential model. This can be understood by noting that the saturation 
density of the $\Lambda=450$\,MeV model is close to the empirical value of 
$n_0 = 0.16$\,fm$^{-3}$, while the other two potentials give saturation densities
closer to $n_0 = 0.17$\,fm$^{-3}$. 

\begin{figure}[t]
\includegraphics[scale=0.45]{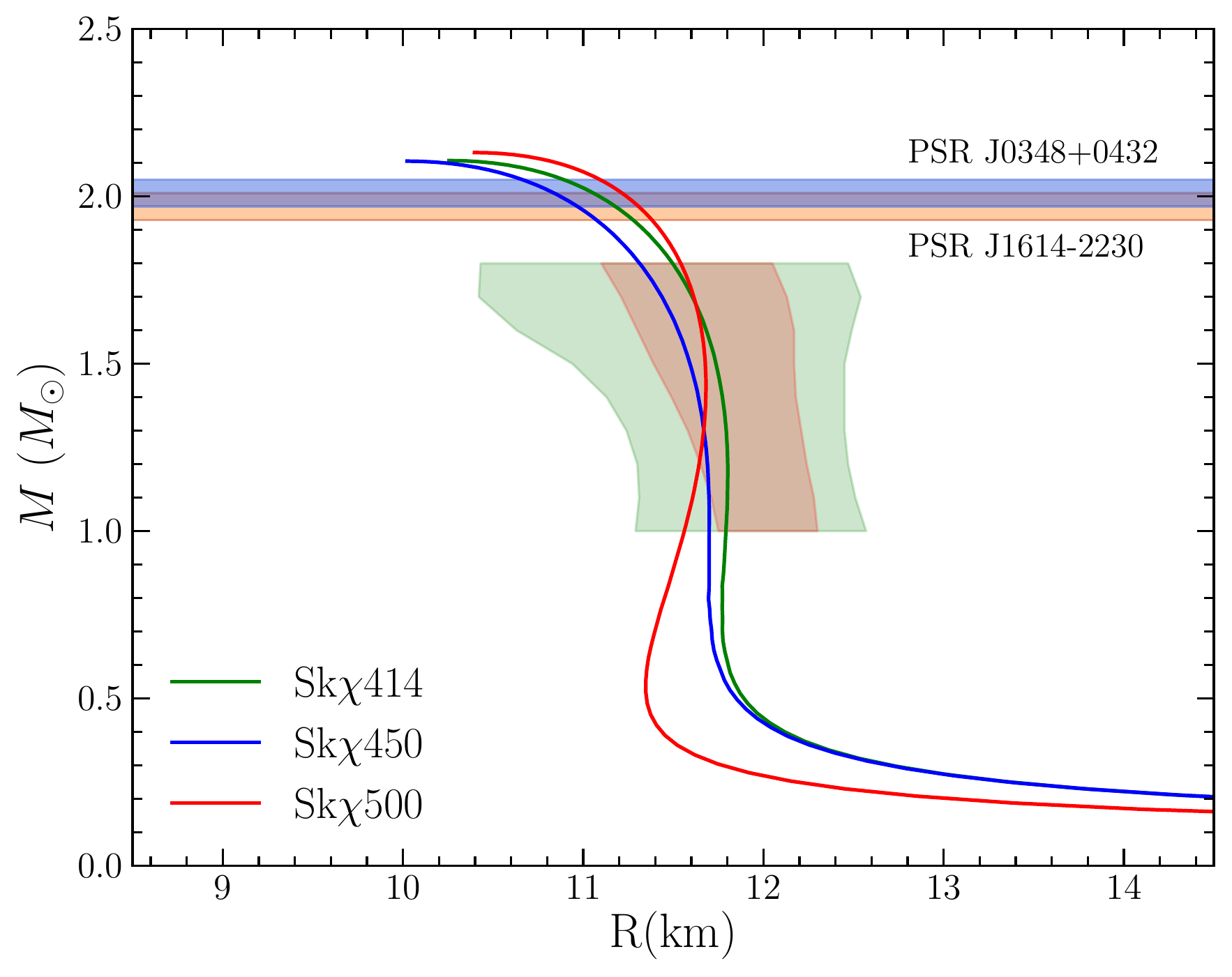}
\caption{(Color online) Mass-radius curves from the Skyrme mean field models
constructed in the present work.}
\label{fig:tov}
\end{figure}

To check the behavior of the Skyrme mean field models in the high-density region
($\rho > 0.4 \mathrm{\,fm}^{-3}$), we solve Tolman-Oppenheimer-Volkov (TOV) equations 
for a static cold neutron star:
\begin{equation}
\begin{aligned}
\frac{dp}{dr} &= - \frac{G(M(r)+4\pi r^{3} p)(\varepsilon + p)}
{r(r-2GM(r))}, \\
\frac{dM}{dr} &= 4\pi \varepsilon r^{2},
\label{eq:tov}
\end{aligned}
\end{equation}
where $r$ is the radial distance from the center,
$M(r)$ is the enclosed mass of a neutron star within $r$, $\varepsilon$ represents
the energy density and $p$ the pressure. Figure \ref{fig:tov} shows the mass 
and radius curves for the three different Skyrme parameter sets.
The central shaded area is a comprehensive estimate of neutron star radii 
from observations of X-ray bursters~\cite{Steiner:2010fz}.
The rectangular bars around $2.0 M_\odot$ represent observational constraints
on the maximum neutron star mass~\cite{Demorest:2010bx,Antoniadis:2013pzd}.
For all three Skyrme parametrizations we see that
the maximum neutron star mass is equal to $2.1 M_\odot$. Therefore, all of the 
parameter sets satisfy the maximum mass constraint and moreover are 
also consistent with the radius constraint.

\section{Core-Crust Boundary}\label{sec:boundary}

\subsection{Asymmetric matter equation of state and nuclear mass tables}
To orient the discussion of the neutron star crust-core transition, we begin 
with a simple model of the crust derived from the BBP~\cite{Baym:1971ax}
formalism. Here the nuclear mass table is used to determine the energy
per nucleon in the crust. When combined with the beta-equilibrium equation 
of state for homogeneous nuclear matter, it is possible to estimate 
the crust-core transition density.
In the BBP formalism, a single nucleus stays at the center of a spherical unit cell
called the ``Wigner-Seitz Cell'' along with a gas of unbound electrons and neutrons.
The total energy density is then given by
\begin{equation}
\vare   = n_N M(A,Z) + n_N W_L + \vare_n(n_n)(1-V_N n_N) + \vare_e(n_e)\,,
\end{equation}
where
$n_N$ is the number density of heavy nuclei with $A$ nucleons ($Z$ protons),
$V_N$ is the volume of a nucleus so that $(1-V_N n_N)$ is the volume fraction
given to neutrons in the Wigner-Seitz cell, 
$W_L$ is the lattice energy arising from the interaction between electrons
and protons in the unit cell, and $n_n$ and $n_e$ are 
the number densities of unbound neutrons and electrons in the cell.
The energy density of neutrons $\vare_n$ is taken from the zero-temperature
neutron matter equation of state from chiral EFT, while the electron energy density 
$\vare_e$ is given by 
\begin{equation}
\vare_e  = \frac{m_e^4}{8\pi^2}
\left[x\sqrt{1+x^2}(1+2x^2) -\ln (x+\sqrt{1+x^2})
\right]\,,
\end{equation}
where $x=k_F^e / m_e$ is the electron Fermi momentum divided by its mass.

Since the Skyrme Hartree-Fock nuclear masses contain the Coulomb energy for 
proton-proton interactions, it is necessary to subtract the Coulomb energy when 
computing the lattice energy. In this work, we consider the exchange Coulomb energy 
from electrons. Thus the total Coulomb lattice energy from electrons and protons is 
given by
\begin{equation}
\begin{aligned}
W_{L+C}^\prime  = & \frac{3}{5}\frac{Z^2e^2}{r_N} 
\left[
      \left(1-\frac{r_N}{r_c}\right)^2
      \left(1 + \frac{r_N}{2r_c}\right) - 1 
\right]  \\
& - \frac{3e^2}{4}\left(\frac{3}{\pi}\right)^{1/3}\frac{n_e^{4/3}}{n_N} \\
= & 
\frac{3}{10}\frac{Z^2e^2}{r_o A^{1/3}}(u - 3u^{1/3})
 - \frac{3}{4}
 \frac{Z^{4/3}e^2}{r_oA^{1/3}}\left(\frac{3}{2\pi}\right)^{2/3}u^{1/3},
\end{aligned}
\end{equation}
where $u$ is the volume fraction of the nucleus in the 
Wigner-Seitz cell, i.e., $V_c = \frac{4\pi}{3}r_c^3 = 1/n_N$ and $u=V_N/V_c$.
The radius of a heavy nucleus in the unit cell is given by $r_N = r_oA^{1/3}$,
where $n_0 = (\frac{4\pi}{3}r_o^3)^{-1} = 0.16~\mathrm{fm}^{-3}$.
Heavy nuclei are therefore assumed to be of uniform density $n_0$. 

\begin{figure}[t]
\includegraphics[scale=0.45]{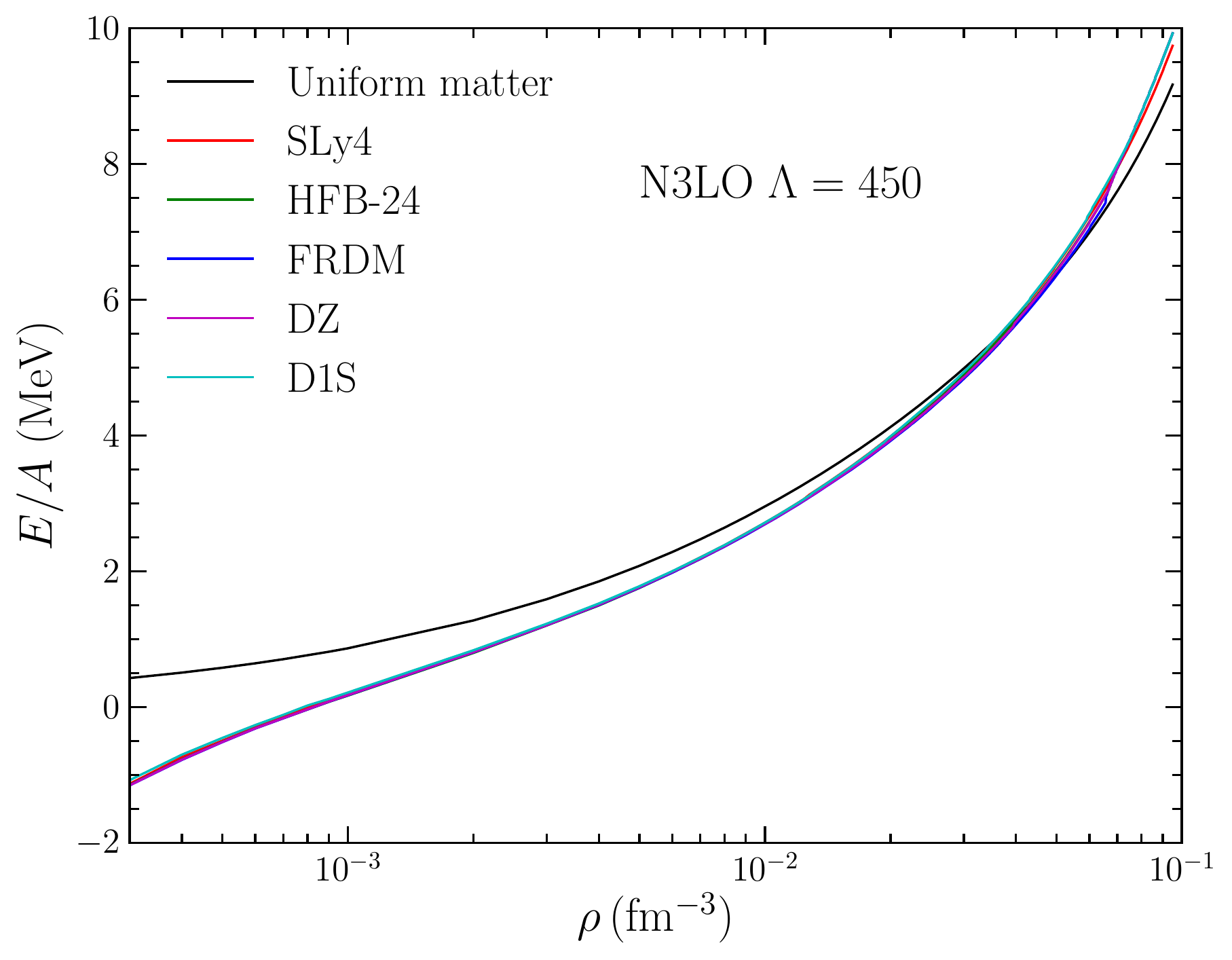}
\caption{(Color online) Energy per nucleon in the neutron star crust for each mass model
together with the n3lo450 beta-equilibrium bulk matter calculation.}
\label{fig:epa_az}
\end{figure}

The approach outlined above gives us a first estimate for the transition density between 
inhomogeneous nuclear matter in the neutron star crust and uniform neutron matter in the core.
The transition to homogeneous matter occurs when the energy density of the Wigner-Seitz
cell containing a heavy nucleus becomes larger than that of homogeneous nuclear matter
in beta-equilibrium.
In Fig.\ \ref{fig:epa_az} we show the energy per baryon in the neutron star crust
using various nuclear mass models together with the bulk matter equation of state
(we take as a representative example that from the 
n3lo450 chiral nuclear potential). All of the nuclear mass models give very similar finite nuclei 
binding energies, but slight differences give rise to crust-core transition densities in the range
$0.035 < \rho_t < 0.052$\,fm$^{-3}$.

In Table \ref{tb:bbp_tr} we show the transition densities using various nuclear mass models
together with the three neutron matter equations of state described in Section \ref{sec:model}.
Note that the Gogny D1S mass model consistently gives the lowest transition
density, which is related to the relatively fast approach to neutron drip in the model.
We observe that the uncertainty in the transition density coming from the choice of nuclear 
mass model is much larger than that from the choice of the bulk matter equation of
state. This is due to the fact that at these relatively low values for the transition density, the 
chiral effective field theory expansion of the nuclear equation of state is well 
converged \cite{Holt:2016pjb}.

Overall, the use of a nuclear mass table together with the bulk matter equation of 
state is a rather crude method to obtain the neutron star crust-core phase boundary. We
will show in more detail below that the model predicts a transition density that is too small,
since each mass table only accounts for the possibility of neutron-rich nuclei in the Wigner-Seitz
cell for which the neutron chemical potential is less than zero. In the inner crust of 
neutron stars, the neutron chemical potential is greater than zero as
neutrons drip out of heavy nuclei to form the free gas of neutrons. Thus,
the mass information of finite nuclei is only useful to describe the neutron star outer crust
\cite{Ruester:2005fm}.

\begin{table}[t]
\caption{The core-crust transition densities in units of fm$^{-3}$ from various nuclear model combinations.
References are for the finite nuclei calculations.}
\begin{tabular}{cccccc}
\hline
\hline
Model     &  Sk$\chi$414         & Sk$\chi$450                 & Sk$\chi$500             &  Ref.\\
\hline
SLy4      & 0.03562          &  0.03556             & 0.03481          & \cite{Stoitsov:2003pd} \\ 
HFB-24    & 0.04256          &  0.04291             & 0.04025          & \cite{Goriely:2013xba} \\
FRDM      & 0.05140          &  0.05196             & 0.04612          & \cite{Moller:2015fba} \\
DZ        & 0.04471          &  0.04512             & 0.04172          & \cite{Duflo:1995ep} \\
D1S       & 0.03505          &  0.03524             & 0.03436          & \cite{Hilaire2007} \\
\hline
\end{tabular}\label{tb:bbp_tr}
\end{table}

\subsection{Compressible Liquid Drop Model}
A more realistic approach to study the neutron star inner crust equation of state is to
utilize the liquid drop model (LDM) in the Wigner-Seitz cell approximation. 
The energy density used to obtain the ground state of inhomogeneous
nuclear matter in the crust of a neutron star can be written as
\begin{equation}
\begin{aligned}
\vare &=  u n_i f_i + \frac{\sigma(x_i)u d}{r_N} 
+ 2\pi(n_i x_i e r_N)^2 u f_d(u)  \\
 &+ (1-u)n_{no}f_{no}\,,
\end{aligned}
\end{equation}
where $u$ is the filling factor (the fraction of space taken up by a heavy nucleus
in the Wigner-Seitz cell),
$n_i$ is the number density of heavy nuclei, $x_i$ is the proton fraction,
$f_i$ represents the volume contribution to the energy per baryon in the heavy nucleus
obtained from the new Skyrme parametrizations,
$\sigma (x_i)$ is the surface tension as a function 
of the proton fraction, $r_N$ is the heavy nucleus radius, 
$n_{no}$ is the density of the unbound neutron gas, $f_{no}$ is the 
energy density of the neutron gas, and $f_d$ is a geometric function describing 
the Coulomb interaction~\cite{Ravenhall:1983uh} for different dimensions $d$.
The surface tension is given explicitly by
\begin{equation}
\sigma(x) = \sigma_0 \frac{2^{\alpha+1} + q }{(1-x)^{-\alpha} + q + x^{-\alpha}},
\label{st}
\end{equation}
where $q$ parametrizes how quickly the surface tension decreases as a function of the 
proton fraction $x$. Larger values of $q$ correspond to more gradual decreases in the 
surface tension for neutron-rich nuclei.
The parameterization of the surface tension in Eq.\ (\ref{st}) avoids the problem of negative 
values that can occur for highly neutron-rich nuclei when a simple quadratic formula 
for the surface tension is used~\cite{Ravenhall:1984ss}. 
The numerical values of $\sigma_0$ and $q$ are fitted to give the lowest root-mean-square 
deviation to known nuclear masses. For the three chiral interactions n3lo414, n3lo450 and 
n3lo500, we find $\sigma_0 = \{1.311,\, 1.186,\, 1.233\}$\,MeV-fm$^{-2}$ and 
$q = \{ 40.362, \,46.748, \, 69.413 \}$,
respectively. In all cases $\alpha=3.4$ is used since it is adequate in describing both isolated 
nuclei and nuclei in dense matter.

\begin{figure}[t]
\includegraphics[scale=0.45]{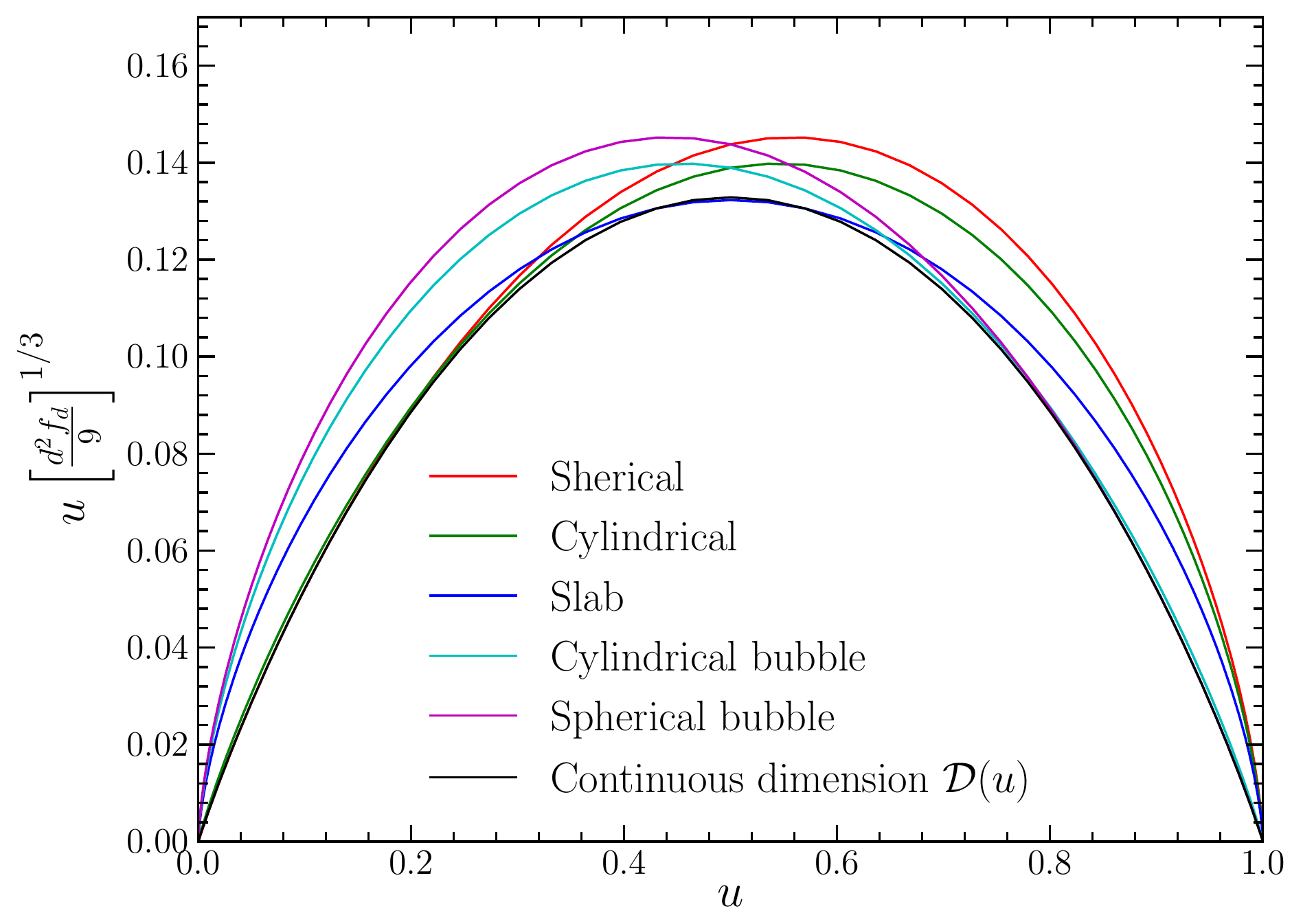}
\caption{(Color online) Shape function $\mathcal{D}(u)$ for discrete dimension and
continuous dimension. The continuous dimension curve always lies below those of the 
discrete geometries.}
\label{fig:shapeu}
\end{figure}

The Coulomb energies for different nuclear geometries (e.g., cylindrical or planar) are 
encoded in the function
\begin{equation}
f_d(u) = \frac{1}{d+2}
\left[ 
\frac{2}{d-2}\left( 1 -\frac{1}{2}du^{1-2/d}\right) + u
\right].
\end{equation}
The case $d=3$ corresponds to spherical shape, $d=2$ to cylindrical shape, and
$d=1$ to slab shape. The equation for spherical bubble geometry can be obtained
with the replacement $u\sigma \rightarrow (1-u)\sigma$ and 
$uf_d(u) \rightarrow (1-u)f_d(1-u)$. 
For a given baryon number density $n$ and
proton fraction $Y_p$, we solve the following equations for the four unknowns 
\{$u$, $n_i$, $x_i$, $n_{no}$\}:
\begin{subequations}
\begin{align}
& \mu_{ni} - \frac{x_i \sigma^{\prime}(x_i) d}{r_N n_i} = \mu_{no}\,, \\
& P_i -2\pi(n_i x_i e r_N)^2 \frac{\partial (uf_d)}{\partial u} = P_{no}\,, \\
& n - un_i - (1-u)n_{no} = 0\,,\\
& nY_p - un_i x_i = 0\,,
\end{align}
\end{subequations}
where $n$ is the total baryon number density in the Wigner-Seitz cell.
From the nuclear virial theorem the surface energy $E_S = \sigma(x_i)u d / r_N$ 
is related to the Coulomb energy $E_C = 2\pi(n_i x_i e r_N)^2 u f_d(u)$ by
$E_S = 2E_C$, which is obtained by setting
$\partial \vare/\partial r_N = 0$. 
This gives \cite{Lattimer:1991nc} the relation $E_S + E_C = \beta \mathcal{D}$, where
$\beta = \left(\frac{243\pi}{2} \right)^{1/3} (n_i x_i e \sigma)^{2/3}$ and
$\mathcal{D}(u) = u \left[\frac{d^2 f_d}{9} \right]^{1/3}$.
If we allow $d$ to be continuous, we can find the shape function $\mathcal{D}$ that
describes all pasta phases with a single formula. 

We adopt the function $\mathcal{D}$ used
in the Lattimer-Swesty EOS~\cite{Lattimer:1991nc}:
\begin{equation}\label{eq:contd}
\mathcal{D}(u) = u(1-u)
\frac{(1-u)f_3^{1/3} + u f_3^{1/3}(1-u)}
{u^2+(1-u)^2 + 0.6u^2(1-u)^2}\,.
\end{equation}
The combined pasta phase model can be implemented if a 
continuous dimension $d$ is allowed. Fig.\ \ref{fig:shapeu}
shows the shape function $\mathcal{D}(u)$ for each discrete dimension (shown
as colored lines) as well as for continuous dimension (black line). 
The latter has the correct behavior as $u\rightarrow 0$ and $u\rightarrow 1$.
It represents the energy state that minimizes the combined Coulomb 
and surface energies.
\begin{figure}[t]
\includegraphics[scale=0.45]{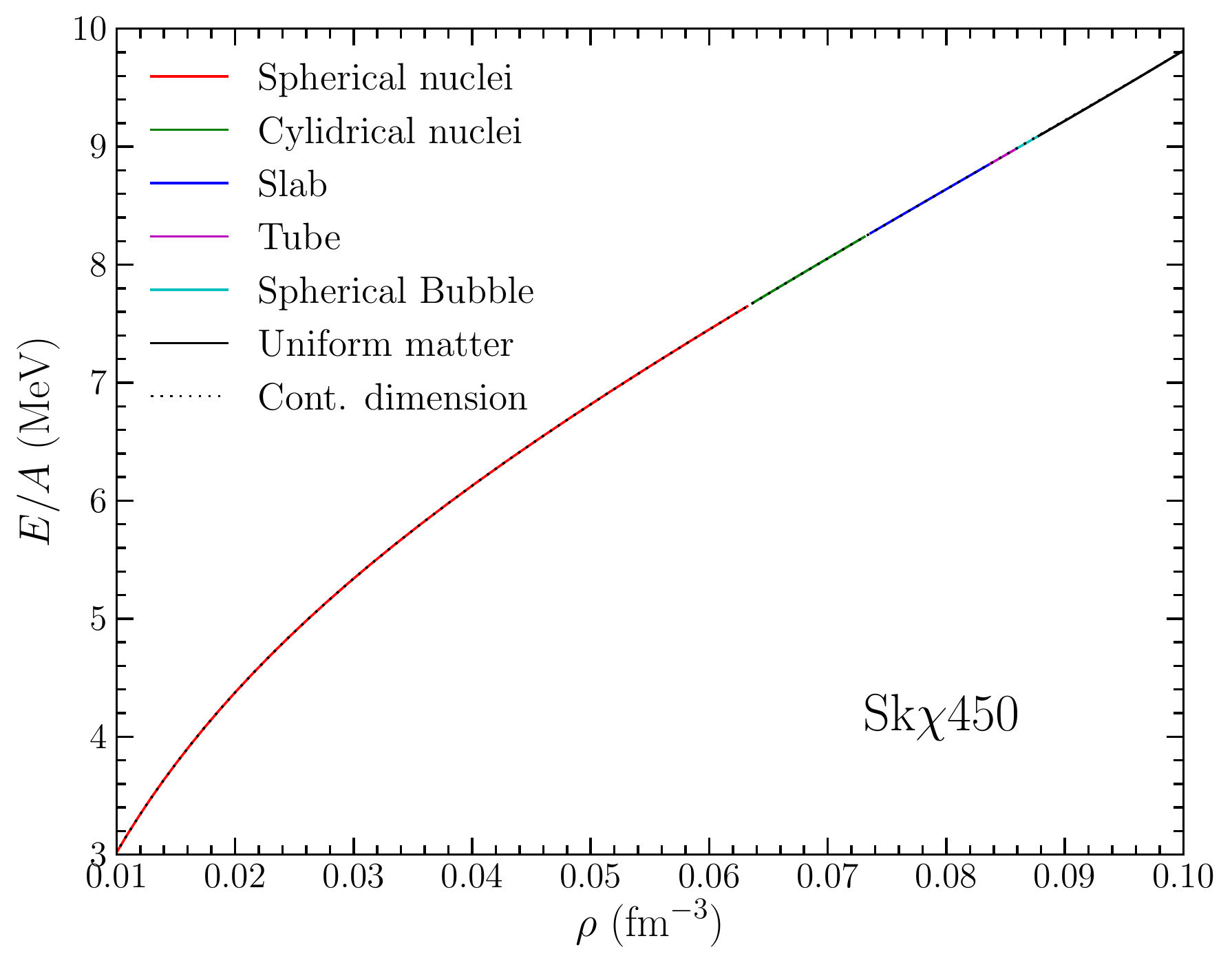}
\caption{(Color online) Energy per nucleon as a function of baryon number density 
in beta-stable nuclear matter employing the liquid drop model with the Sk$\chi$450
Skyrme mean field model.}
\label{fig:epaldm}
\end{figure}

Note that the dimension of the lowest energy state will be determined by the
volume fraction of dense matter in the Wigner-Seitz cell. 
The crossing points for each dimension are independent of the equation of
state and occur at the values
$u = \{0.21525, 0.35499, 0.64501, 0.78475\}$ for the 
\{3D-2D,  2D-1D, 1D-2DB, 2DB-3DB\} transitions. For instance,
if the volume fraction of dense matter is 0.4, then the lowest energy state is
the slab phase. 

In Fig.\ \ref{fig:epaldm} we show the energy per baryon in the geometric configuration
with the lowest energy, including also the beta-equilibrium condition.
As the density increases the lowest energy state proceeds through $d = 3, 2, 1, 2b, 3b,$ and 
finally to uniform matter. By ``2b'' and ``3b'' we denote the two-dimensional and three-dimension
bubble geometries. The solution found by employing a continuous dimension correctly represents 
the lowest energy state.
\begin{figure}[t]
\includegraphics[scale=0.45]{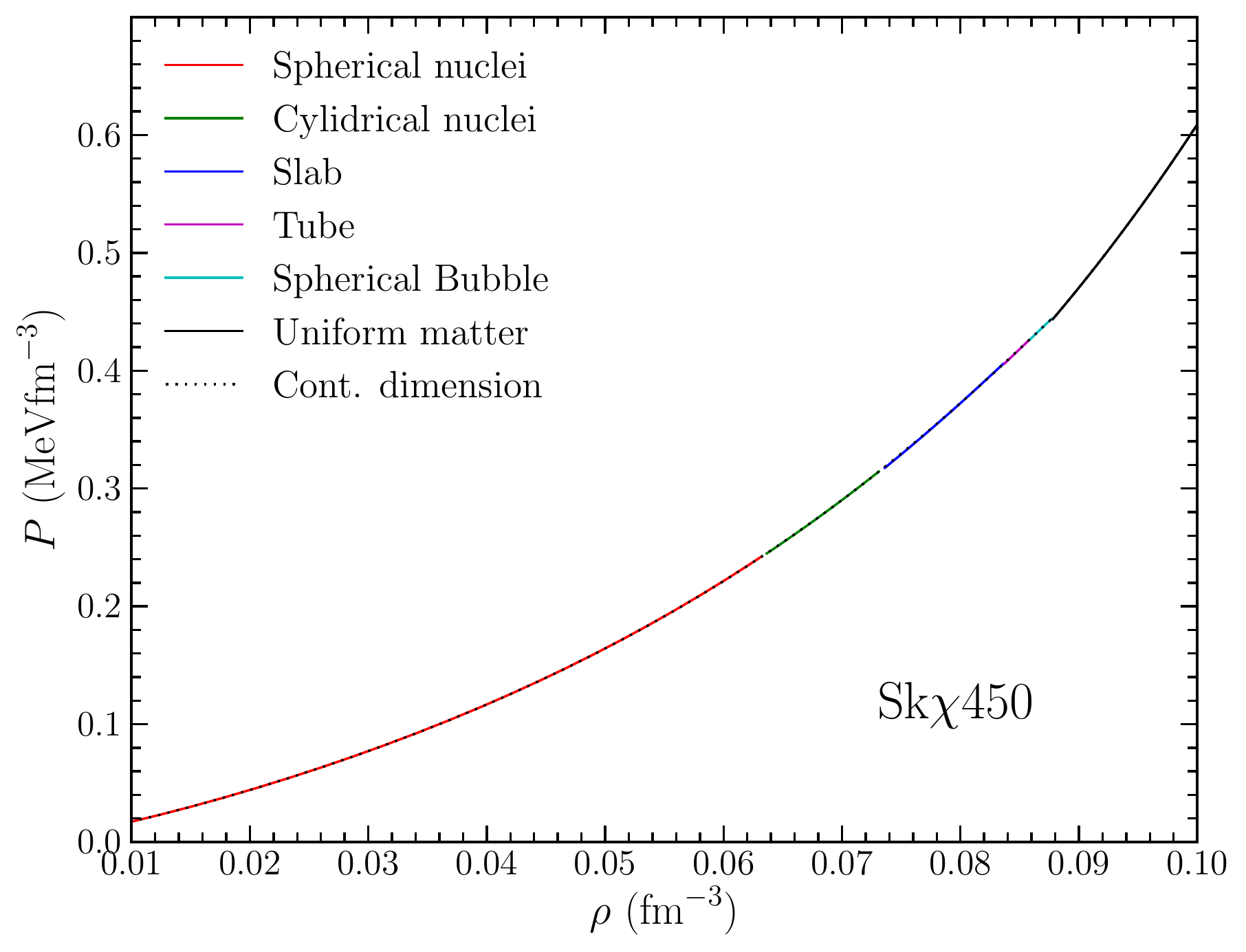}
\caption{(Color online) Pressure as a function of baryon number density in beta-stable 
nuclear matter employing the liquid drop model with the Sk$\chi$450
Skyrme mean field model. At each transition density the pressure is almost continuous 
in the case of the LDM approach.}
\label{fig:ldmpre}
\end{figure}
The first derivative of $E/A$ with respect to the baryon number density, namely the pressure,
is shown in Fig.\ \ref{fig:ldmpre}. The pressure at each transition density is essentially
continuous in the LDM formalism. The continuous dimension LDM also gives
the correct numerical values compared with the discrete dimension calculation in the LDM. 

In Table \ref{tb:ldmtr} we show the phase transition densities to different nuclear
pasta geometries in the neutron star inner crust. 
We see that the different Sk$\chi$ mean field models predict similar transition densities
for each of the phases,
with uncertainties less than $0.006$\,fm$^{-3}$.
\begin{table}[h]
\caption{Transition densities (in unit of fm$^{-3}$) between different geometries in the 
neutron star inner crust using the LDM method.}
\label{tb:ldmtr}
\begin{tabular}{ccccc}
\hline
\hline
           & ~Sk$\chi$414~ & ~Sk$\chi$450~    & ~Sk$\chi$500~ \\    
\hline
3DN-2DN    &  0.0665           & 0.0634               & 0.0656 \\
2DN-1DN    &  0.0766           & 0.0736               & 0.0782 \\
1DN-2DB    &  0.0864           & 0.0837               & 0.0895 \\
2DB-3DB    &  0.0884           & 0.0859               & 0.0918 \\
3DB-Uni.   &  0.0901           & 0.0878               & 0.0940 \\
\hline
\end{tabular}
\end{table}

\begin{figure}[t]
\includegraphics[scale=0.45]{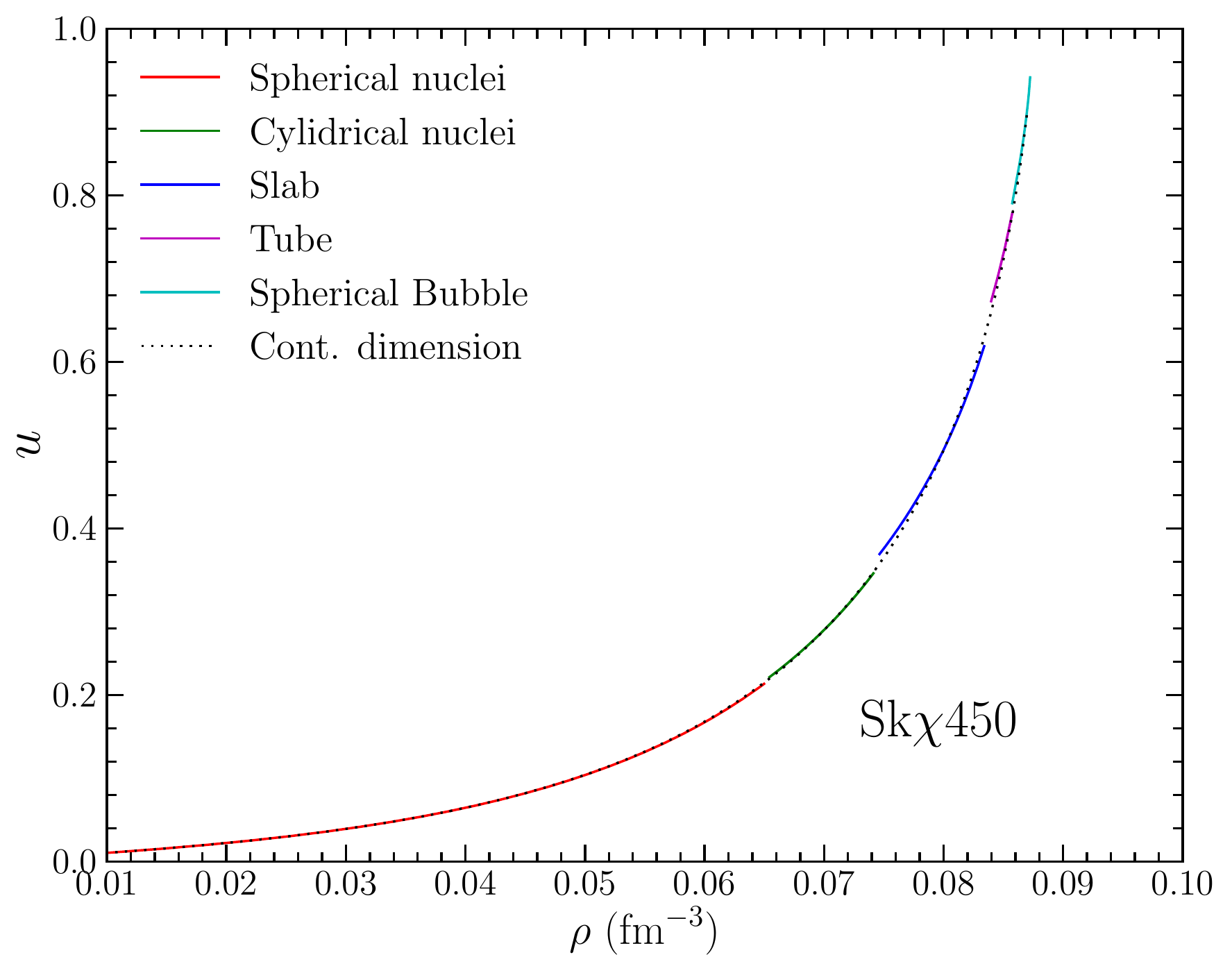}
\caption{(Color online) Volume fraction of the dense phase in the Wigner-Seitz cell.
The volume fraction indicates which dimension is the ground state for a given
baryon number density.}
\label{fig:volu}
\end{figure}

In Fig.\ \ref{fig:volu} we show the volume fraction of dense matter
in the Wigner-Seitz cell for each discrete dimension and continuous 
dimension calculation. The volume fractions for the lowest energy states 
are in the correct regions as expected.
Therefore, the volume fraction of the dense phase in the Wigner-Seitz cell 
at each dimension can be used to identify the ground state dimension 
among the different pasta phases.
The continuous dimension approach provides a reliable way to construct 
the nuclear equation of state in the pasta phase analytically.
This also indicates that the supernova EOS table~\cite{Lattimer:1991nc} using 
the continuous dimension is a valid numerical method that does not
destroy the continuity in pressure at each transition density.

\subsection{Thomas-Fermi Approximation}
In the Thomas-Fermi~(TF) approximation, the number density and kinetic momentum density 
are given by
\begin{equation}
\rho_t =  \frac{1}{4\pi^2} \int_{0}^{\infty} f_t d^3p\,,\quad
\tau_{t} = \frac{1}{4\pi^2} \int_{0}^{\infty}f_t p^2 d^3p
\end{equation}
where $t$ is the type of nucleon and $f_t$ is the Fermi occupation function:
\begin{equation}
f_t = \frac{1}{1 + \exp{ \left( \frac{\vare_{t} - \mu_t}{T} \right) } }\,,
\end{equation}
where $\vare_t$ is the single particle energy for protons or neutrons
and $\mu_t$ is the chemical potential for each species.
At $T=0$~MeV this equation simply gives $\tau_t = \frac{3}{5}(3\pi^2)^{2/3}\rho_t^{5/3} $.
In the crust of neutron stars, 
the density profile of inhomogeneous nuclear matter can be parametrized~\cite{Oyamatsu:1993zz} as
\begin{equation}
n_t(r) = \begin{cases}
(n_{ti} - n_{to})
 \left[1 -\left(\frac{r}{R_t}\right)^{\alpha_t}\right]^3 + n_{to} & 
 \text{if} \quad r< R_t \,, \\
 n_{to}  & \text{if} \quad r \ge R_t\,. 
\end{cases}
\end{equation}
When $\mu_n >0 $, $n_{no} \ne 0$. Thus $n_{no}$ represents the density of the unbound 
neutron gas.
Depending on the density, all parameters ($n_{ti}$, $n_{to}$, $r_t$, $R_t$, $\alpha_t$) are
to be obtained numerically from the minimization of the total energy:
\begin{equation}
\begin{aligned}
E= \int\Bigl[ &
\mc{H}(n_n, n_p) + m_n n_n + m_p n_p + \mc{E}_{el}(n_e) \\
& 
+ \mc{E}_{Coul}(n_p,n_e) + \mc{E}_{ex}(n_p,n_e) 
\Bigr] \,d \mb{r}\,,
\end{aligned}
\end{equation}
where the Hamiltonian $\mc{H}$ is given by
\begin{equation}
\mc{H}(n_n, n_p )
= \frac{1}{2m_n}\tau_n + \frac{1}{2m_p}\tau_p
+ V_{NN}(n_n, n_p)\,.
\end{equation}
We use for $V_{NN}$ the non-relativistic Skyrme force models obtained in this work.
In the crust of neutron stars, the electrons are distributed 
uniformly, so we assume a constant electron density. The Coulomb energy
is given by
\begin{equation}
\mc{E}_{Coul}(n_p, n_e) 
 = \frac{1}{2}\Bigl[ n_p(r) -n_e \Bigr]\Bigl[ V_p(r) - V_e(r)\Bigr]\,.
\end{equation}
The Coulomb potentials for protons and electrons are given by
\begin{equation}
V_p(\mbr) = \int \frac{e^2\, n_p(\mbr^\prime)}{|\mbr -\mbrp|}\,d\mbrp\,, \quad
V_e(\mbr) = \int \frac{e^2\, n_e}{|\mbr -\mbrp|}\,d\mbrp\,
\end{equation}
and the Coulomb exchange energy is given as
\begin{equation}\label{eq:ex}
\mc{E}_{ex} = - \frac{3}{4} \left( \frac{3}{\pi} \right)^{1/3}e^2 
\left[ n_p^{4/3}(\mbr) + n_e^{4/3}\right]\,.
\end{equation}
\begin{figure}[t]
\includegraphics[scale=0.45]{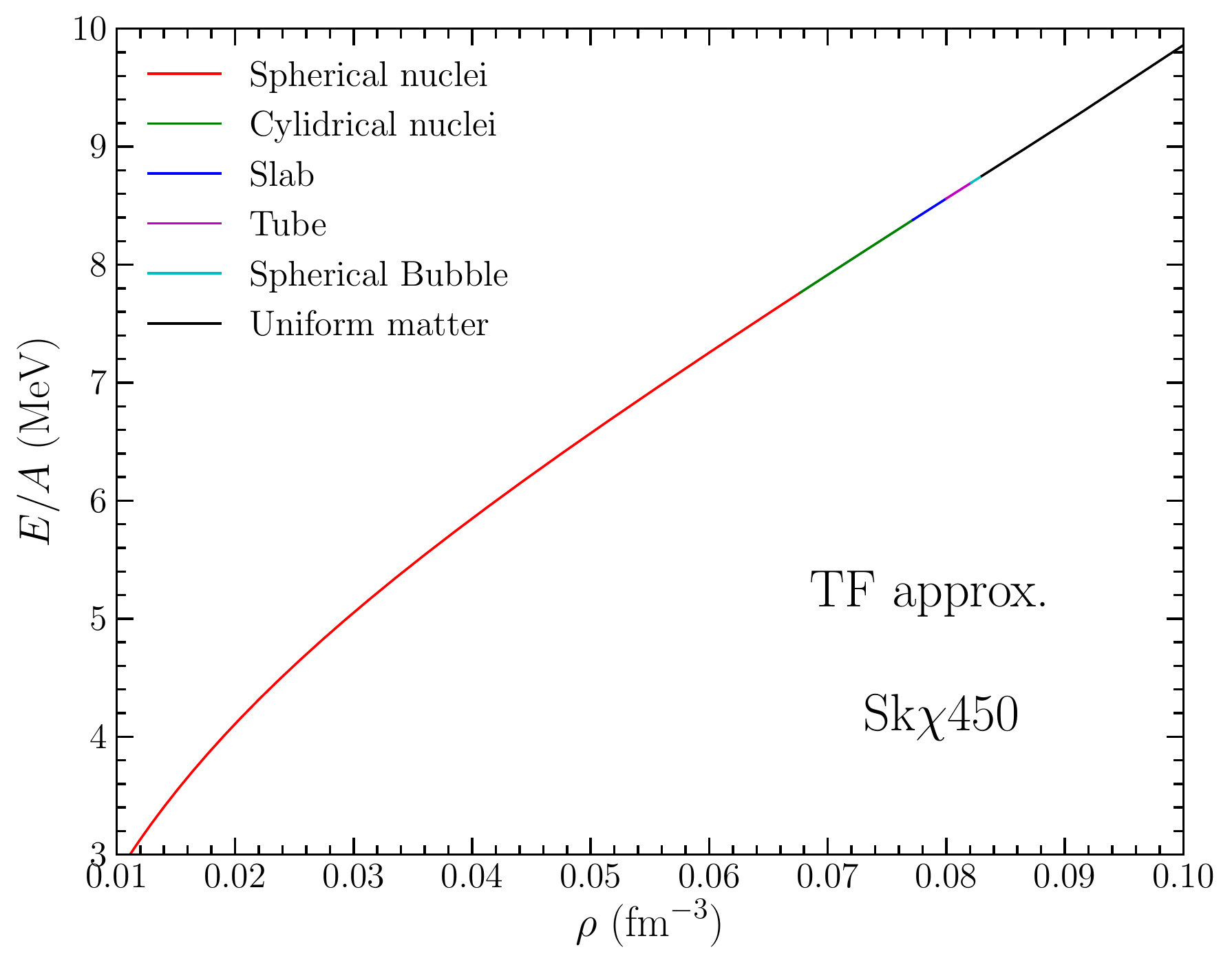}
\caption{(Color online) Energy per baryon using the TF approximation with the Sk$\chi$450 Skyrme fit model.}
\label{fig:tfepa}
\end{figure}
The nuclear pasta phases require
Coulomb interaction formulas for different dimensions \cite{Sharma:2015bna}:
\begin{equation}
\begin{aligned}
& \text{Spherical :}  \\
V_p(r) & = 4\pi e^2 \biggl[\frac{1}{r}\int_0^r r^{\prime 2} \rho_p(r^\prime) dr^\prime 
+ \int_r^{R_c} r^\prime \rho_p(r^\prime) dr^\prime \biggr] \,, \\
V_e(r) & = 2\pi e^2 n_e \left[ R_c^2 - \frac{1}{3}r^2 \right]\,.
\end{aligned}
\end{equation}
\begin{equation}
\begin{aligned}
& \text{Cylindrical :}\\
V_p(r) & = -4\pi e^2 \biggl[\ln(r)\int_0^r r^{\prime} \rho_p(r^\prime) dr^\prime 
+ \int_r^{R_c} r^\prime \ln r^\prime \rho_p(r^\prime) dr^\prime \biggr] \,, \\
V_e(r) & = \pi e^2 n_e R_c^2\left[ 1- \frac{r^2}{R_c^2} - 2 \ln R_c \right]\,.
\end{aligned}
\end{equation}
\begin{equation}
\begin{aligned}
& \text{Slab :} \\
V_p(z) & = -4\pi e^2 \left[z \int_0^z\rho_p(z^\prime) dz^\prime
+ \int_z^{R_c} z^\prime  \rho_p(z^\prime) dz^\prime \right] \,, \\
V_e(z) & = -2 \pi e^2 n_e (z^2 + R_c^2)\,.
\end{aligned}
\end{equation}

Fig.\ \ref{fig:tfepa} shows the energy per baryon for beta-equilibrated neutron star
matter obtained in the TF approximation
using the Skyrme parametrization Sk$\chi$450 developed in the present work. As in the 
case of the LDM model, the ground-state geometry for increasing density proceeds 
through the sequence \{spherical, cylindrical, slab, cylindrical hole, spherical hole, uniform matter\} 
in this order. Each new geometry spans smaller and smaller ranges of densities, and the transition
density to the homogeneous phase occurs at $n_c = 0.084$\,fm$^{-3}$.
\begin{figure}[t]
\includegraphics[scale=0.45]{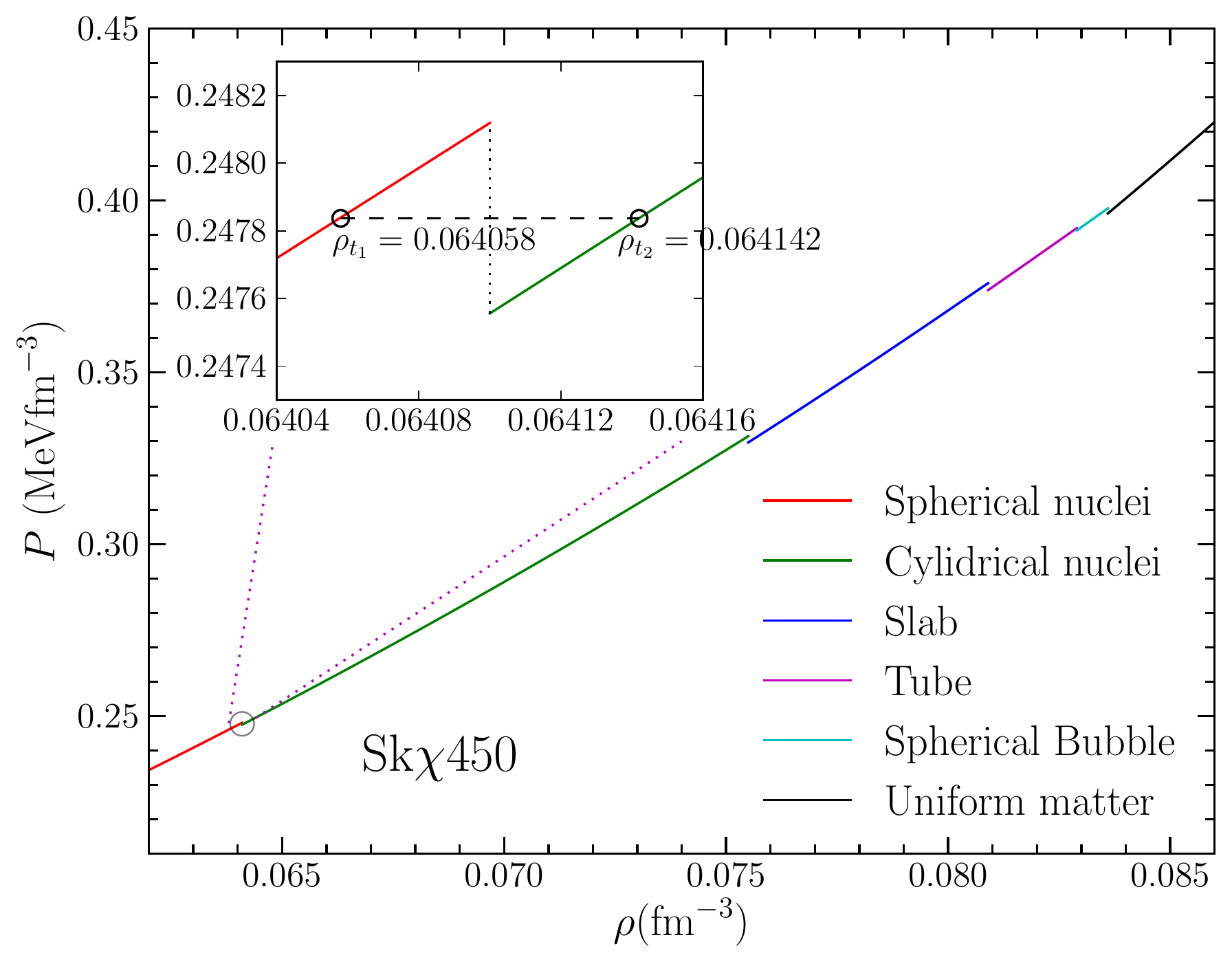}
\caption{(Color online) Pressure vs.\ baryon number density using in the TF approximation. 
A discontinuity in the pressure occurs
at the shape transition densities, but the discontinuity region (shown in the inset) is very narrow. }
\label{fig:tfpre}
\end{figure}

The ground state pressure as a function of density employing the same interaction
model is shown in Fig.\ \ref{fig:tfpre}.
Unlike the LDM approach, the Thomas-Fermi approximation results in a small discontinuity
in the pressure at the interface between each phase when we only compare the energy per 
baryon to find the ground state of the phase. 
This is caused by the intrinsic discontinuity in the expressions for the Coulomb energy
in the different geometries. The LDM approach enables us to investigate the structure
of the pasta phase with fewer parameters, so the pressure discontinuity or proton
fraction discontinuity can be small. On the other hand, the more realistic TF method
can be done in the space discretization. This means that the discontinuity in the pressure
is a natural phenomenon in the case of phase transformation in the TF approximation. 
When the Maxwell construction is employed, the interval of the density in the
coexistence region is so small ($\Delta \rho = 0.0001$~fm$^{-3}$) that the microscopic
structure of the neutron star barely changes. As an example, the two densities of mixed state
for spherical shape and cylindrical shape are $\rho_{t_1}=0.06406$ and 
$\rho_{t_2}=0.06414$~fm$^{-3}$. 
\begin{figure}[t]
\includegraphics[scale=0.45]{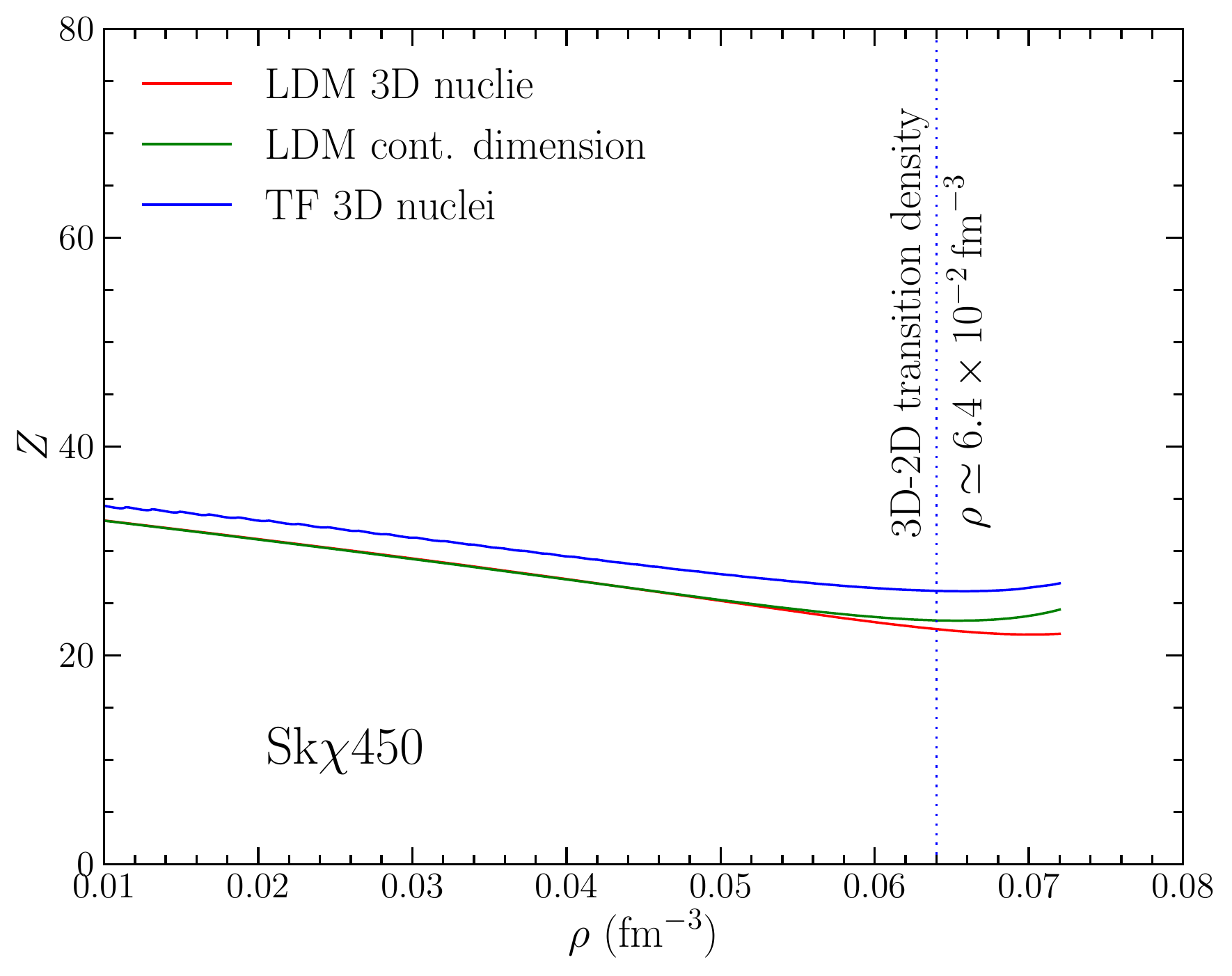}
\caption{(Color online) Atomic number of heavy nucleus in the Wigner Seitz cell.
The dotted line around $\rho=0.064$fm$^{-3}$
indicates the transition between 3D nuclei and 2D nuclei.}
\label{fig:az}
\end{figure}

The choice of LDM vs.\ TF model also gives rise to 
differences in nuclear composition. Fig.\ \ref{fig:az} shows
the atomic number of heavy nuclei in the crust of neutron stars. 
The dotted line indicates the $3D-2D$ phase transition density, which is nearly
independent of whether we employ the LDM or the TF model. The
atomic number is consistently larger in the TF approximation, differing from the 
LDM atomic number by roughly two up to the transition to cylindrical geometry.
The atomic number in continuous dimension over the $3D-2D$ phase transition density
represents the average atomic number in the unit cell. 
It is not a physical quantity in
the crust. Above the $3D-2D$ phase transition density, the TF model gives
a larger atomic number since the Wigner-Seitz cell decreases as the total baryon
density increases (which means the distance between nuclei decreases)
and total number of protons and neutrons increases in the spherical cell.
\begin{figure}[t]
\includegraphics[scale=0.45]{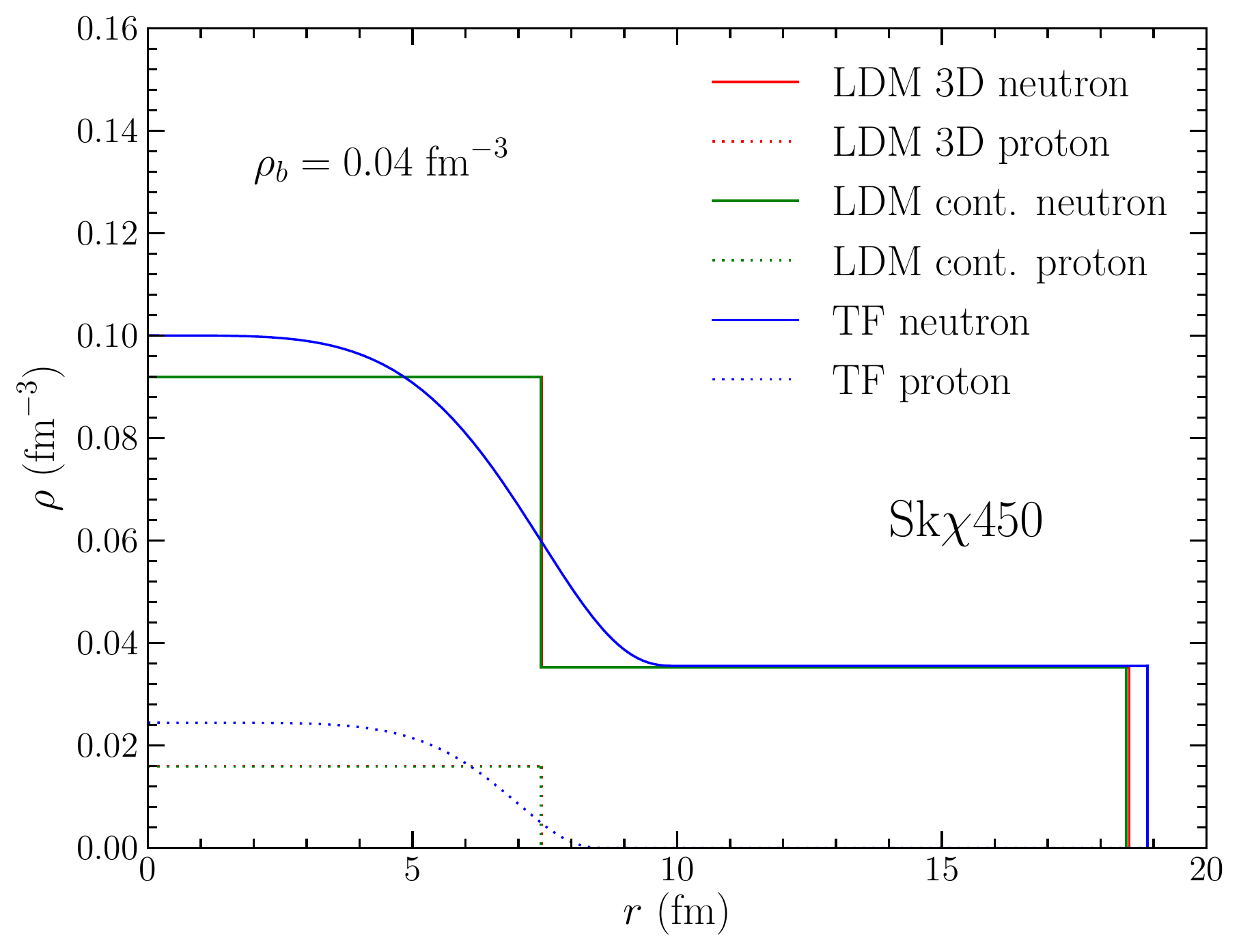}
\caption{(Color online) Neutron and proton density profiles using three different numerical 
methods with the Sk$\chi450$ Skyrme mean field model.}
\label{fig:prof}
\end{figure}

Fig.\ \ref{fig:prof} shows the neutron and proton density profiles
in each numerical calculation with the Sk$\chi450$ interaction.
Even if the central densities of protons and neutrons are different in the 
LDM and TM model, the neutron densities outside the nucleus are nearly 
the same. This indicates that the density profile is the problem to be solved 
in order to understand the coexistence of dense and dilute matter.
Whatever numerical method is used, the density of the unbound gas of neutrons 
should be the same under identical physical conditions.

\begin{table}[h]
\caption{Pasta phase transition densities (in units of fm$^{-3}$) using the TF method. The
numbers in parentheses represent the transition densities with the exchange Coulomb interaction
included.}
\label{tb:tfrhot}
\begin{tabular}{ccccc}
\hline
\hline
                          &  ~~Sk$\chi414$~~  & ~~Sk$\chi450$~~      & ~~Sk$\chi500$~~ \\
\hline
\multirow{2}{*}{3DN-2DN}  &  0.0681           & 0.0641               & 0.0626 \\
                          &  (0.0682)         & (0.0642)             & (0.0627)\\
\multirow{2}{*}{2DN-1DN}  &  0.0791           & 0.0755               & 0.0790 \\
                          &  (0.0795)         & (0.0758)             & (0.0793)\\
\multirow{2}{*}{1DN-2DB}  &  0.0830           & 0.0809               & 0.0865\\
                          &  (0.0838)         & (0.816)              & (0.0869)\\
\multirow{2}{*}{2DB-3DB}  &  0.0852           & 0.0830               & 0.0885\\
                          &  (0.0862)         & (0.0836)             & (0.0891)\\
\multirow{2}{*}{3DB-Uni.} &  0.0860           & 0.0835               & 0.0894\\
                          &  (0.0869)         & (0.0843)             & (0.0894)\\
\hline
\end{tabular}
\end{table}

Table \ref{tb:tfrhot} shows the transition density at each phase boundary.
The transition density for uniform matter is highly correlated with
the saturation density. If the saturation density is greater (as is the case for
the Sk$\chi$414 and Sk$\chi$500 Skyrme interactions), uniform 
nuclear matter is formed at a higher density.
The numbers in parentheses indicate the transition
density when we include the exchange Coulomb interaction in the numerical
calculation.
The exchange Coulomb interaction in Eq.\ \eqref{eq:ex} gives a negative contribution to
the total energy and therefore its presence tends to delay the transitions to higher 
densities. However, the effects are nearly negligible. 

\subsection{Thermodynamic instability}
In neutron stars, the phase transition from uniform nuclear matter to inhomogeneous nuclear matter
takes place when matter begins to exhibit an instability to density fluctuations. 
Baym et al.~\cite{Baym:1971ax} show that the matter is stable when the following relationship is maintained:
\begin{equation}
v_0 + 2(4\pi e^2 \beta)^{1/2} - \beta k_{TF}^2>0\,,
\end{equation}
where 
\begin{equation}
v_0 = \frac{\pt \mu_p}{\pt \rho_p} - \frac{(\pt \mup/\pt \rho_n)^2}{\pt \mun/ \pt \rho_n}\,,
\end{equation}
\begin{equation}
\beta = 2(Q_{pp} + 2Q_{np}\zeta + Q_{nn}\zeta^2 )\,,\quad 
\zeta = - \frac{\pt \mu_p /\pt \rho_n}{\pt \mun/\pt \rhon}\,,
\end{equation}
and $k_{TF}$ is the Thomas-Fermi wave number,
\begin{equation}
k_{TF}^2 = \frac{4e^2}{\pi}k_{e}^2\,,
\quad k_e = (3\pi^2 \rhop)^{1/3}\,.
\end{equation}
\begin{figure}[t]
\includegraphics[scale=0.45]{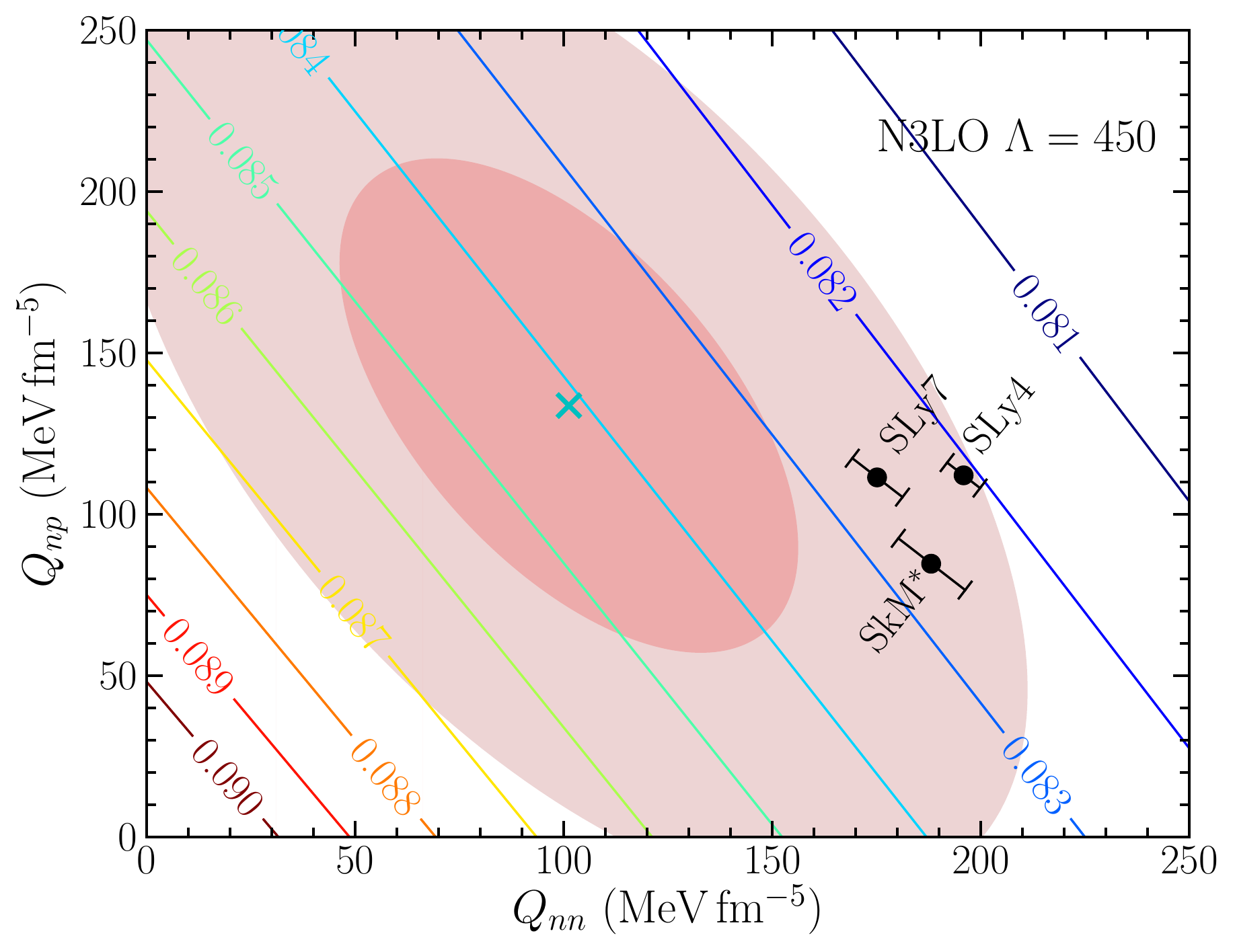}
\caption{(Color online) Transition density contour plot for the core-crust boundary obtained from
thermodynamic instability. The individual points are taken from the modified Skyrme
interactions obtained in Ref.\ \cite{Buraczynski:2016jia}.}
\label{fig:qq}
\end{figure}

In Skyrme models, $Q_{nn}$ and $Q_{np}$ are given by
\begin{equation}
\begin{aligned}
 Q_{nn} = Q_{pp}  = & \frac{3}{16}\left[ t_1(1-x_1) - t_2(1+x_2) \right] \,, \\
 Q_{np} = Q_{pn}  = & \frac{1}{16}\left[ 3t_1 (2 + x_1) - t_2(2 + x_2)\right]\,.
\end{aligned} 
\end{equation}
For the three Skyrme parametrizations developed in this work, 
$Q_{nn}$ and $Q_{np}$ are given by
$Q_{nn}=\{ 107.297, 105.458, 106.901\}$\,MeV-fm$^{-5}$ and 
$Q_{np}=\{119.833, 94.759, 119.641\}$\,MeV-fm$^{-5}$ for Sk$\chi414$, Sk$\chi450$, and Sk$\chi500$
respectively. A more conservative uncertainty estimate is obtained by considering
a wider set of 31 Skyrme models whose equations of state are similar to that from chiral 
effective field theory. 
Fig.\ \ref{fig:qq} shows the resulting confidence contour of $Q_{nn}$ and $Q_{np}$, 
with the symbol `$\mathbf{x}$' at the center of the ellipse representing
the average values. In these calculations the proton and neutron chemical potentials 
in homogeneous matter are taken from the microscopic equation of state computed from
the $\Lambda = 450$\,MeV chiral nuclear potential. The three individual points labeled
``SLy7'', ``SLy4'', and ``SkM*'' come from the modified isovector gradient coupling
strengths deduced in recent quantum Monte Carlo studies \cite{Buraczynski:2016jia}.
Fig.\ \ref{fig:qq} indicates that the density for the core-crust boundary is 
between $\rho=0.082$~fm$^{-3}$
and $\rho =0.087$~fm$^{-3}$.
We infer from the contour plot that the core-crust transition density is proportional
to the sum of $Q_{nn} + k Q_{np}$. We propose an empirical formula for 
the core-crust density with $Q_{nn}$ and $Q_{np}$:
\begin{equation}
\label{eq:rtform}
\rho_{t} \simeq \rho_{t_1} + \alpha  Q_{nn} + \beta Q_{np},
\end{equation}
which indicates that
$Q_{nn}$ and $Q_{np}$ will directly determine the core-crust density.
\begin{table}[h]
\caption{Numerical values for the parameters in Eq.\ \eqref{eq:rtform}.}
\begin{tabular}{ll}
\hline\hline
$\rho_{t_1}$ (fm$^{-3}$)                        & ~~~~$9.103\times 10^{-2} \pm 7.065\times 10^{-4}$   \\
\hline
$\alpha$ ($\mathrm{MeV}^{-1}\mathrm{fm}^{2}$)   & ~~$-3.088\times 10^{-5} \pm 5.257\times 10^{-7}$ \\
\hline
$\beta$  ($\mathrm{MeV}^{-1}\mathrm{fm}^{2}$)   & ~~$-1.891\times 10^{-5} \pm 1.010\times 10^{-6}$ \\
\hline 
\end{tabular}
\end{table} 

\section{Conclusion}\label{sec:con}
We have studied the composition and structure of neutron star crusts by comparing
the energy densities for different pasta phases using both the liquid drop 
model and the Thomas-Fermi model. 
The results are based on a new set of extended Skyrme parametrizations 
derived in the present work that fit the
bulk isospin-asymmetric nuclear matter equation of state from
$\chi$EFT and the binding energies of doubly-magic nuclei.
The neutron star maximum masses obtained from these Skyrme parametrizations 
are consistent with observations of 
$2.0 M_\odot$ neutron stars.

From the LDM and TF calculations, the crust-core transition density is strongly 
correlated with the saturation density of symmetric nuclear matter. For this reason
the extended Skyrme parametrization Sk$\chi$450, which reproduces well both the 
empirical saturation energy and density, is expected to provide the most 
reliable prediction for the crust-core interface density. The predicted pressure at the
phase boundaries between different pasta geometries is smooth in the LDM but
exhibits small discontinuities in the TF approximation.
We have studied as well a continuous-dimension LDM that treats the pasta phases 
as a function of the dense matter volume fraction in the Wigner-Seitz cell.
All three methods give a core-crust boundary density around
half saturation density, $\rho_t = 0.084 $fm$^{-3}$. 

Compared to previous works \cite{hebeler13aj,tews17}, we analyzed
the theoretical uncertainties in the core transition density of neutron stars by varying the gradient terms
$Q_{nn}$ and $Q_{np}$. We find that the transition density has a two-dimensional
correlation with the $Q$'s. Low values of these gradient term
coupling strengths result in an increase in the transition density from the crust to core,
which increases the volume of the neutron star crust. 
The uncertainty in $Q_{nn}$ and $Q_{np}$ can be reduced by
microscopic calculations of the static density response function 
using $\chi$EFT in many-body perturbation theory or quantum Monte Carlo simulations. 
A more accurate determination of $Q_{nn}$ and $Q_{np}$ will therefore play an
important role for improving energy density functionals and to
more accurately predict the density at a neutron star's core-crust boundary.

We find that nuclear pasta exists within the density range between $\rho=0.065$~fm$^{-3}$
and $0.090$~fm$^{-3}$. Macroscopically it exists within a 100~m thickness 
in the inner crust of a neutron star with 1.4$\msun$. The spherical hole phase
exists within the density range of $\Delta \rho=0.002$~fm$^{-3}$ at most.
This means that spherical holes exist only within a $\Delta R=5$~m range in
neutron stars, which might be destroyed in fast rotating neutron stars
because of tidal deformation.
Our results are similar to the previous works of Oyamatsu~\cite{Oyamatsu:1993zz}
and Sharma et al.~\cite{Sharma:2015bna}, who employed phenomenological models
with equations of state similar to the predictions from $\chi$EFT.


\bibliographystyle{apsrev4-1}

%

\end{document}